\documentclass[a4paper,fleqn,usenatbib]{mnras}

\usepackage[T1]{fontenc}
\usepackage{ae,aecompl}
\usepackage{relsize} 
\usepackage{color,soul}
\usepackage{subfig}

\usepackage{graphicx}
\usepackage{amsmath}
\usepackage{amssymb}
\usepackage{bm}
\usepackage{footnote}
\usepackage{bigints}

\usepackage{newtxtext,newtxmath}

\title[External photoevaporation \& planet formation]{\vspace{-2mm}The growth and migration of massive planets under the influence of external photoevaporation\vspace{-3mm}}

\author[A.~J.~Winter et al.]{Andrew~J.~Winter$^{1}$\thanks{andrew.winter@uni-heidelberg.de}, Thomas J. Haworth$^{2}$, Gavin A. L. Coleman$^{2}$ and Sergei Nayakshin$^{3}$\\
$^{1}$Zentrum f\"{u}r Astronomie, Heidelberg University, Albert Ueberle Str. 2, 69120 Heidelberg, Germany \\
$^{2}$Astronomy Unit, Queen Mary University of London, Mile End Road, London, E1 4NS, UK\\
$^{3}$School of Physics and Astronomy, University of Leicester, Leicester, LE1 7RH, UK
\vspace{-3mm}
}

\date{Accepted X{\sevensize xxxx} XX. Received X{\sevensize xxxx} XX; in original form March 7 2022}\vspace{-2mm}

\pubyear{2021}

\begin{document}
\label{firstpage}
\pagerange{\pageref{firstpage}--\pageref{lastpage}}
\maketitle

\begin{abstract}
The formation of gas giant planets must occur during the first few Myr of a star's lifetime, when the protoplanetary disc still contains sufficient gas to be accreted onto the planetary core. The majority of protoplanetary discs are exposed to strong ultraviolet irradiation from nearby massive stars, which drives winds and depletes the mass budget for planet formation. It remains unclear to what degree external photoevaporation affects the formation of massive planets. In this work, we present a simple one dimensional model for the growth and migration of a massive planet under the influence of external FUV fields. {We find that even moderate FUV fluxes $F_\mathrm{FUV}\gtrsim 100 \, G_0$ have a strong influence on planet mass and migration. By decreasing the local surface density and shutting off accretion onto the planet, external irradiation suppresses planet masses and halts migration early. The distribution of typical stellar birth environments can therefore produce an anti-correlation between semi-major axis and planet mass, which may explain the apparent decrease in planet occurrence rates at orbital periods $P_\mathrm{orb}\gtrsim 10^3$~days. Even moderate fluxes $F_\mathrm{FUV}$ strongly {suppress} giant planet formation and inward migration for any initial semi-major axis if the stellar host mass $M_*\lesssim 0.5\, M_\odot$, consistent with findings that massive planet occurrence is much lower around such stars.} The outcomes of our prescription for external disc depletion show significant differences to the current approximation adopted in state-of-the-art population synthesis models, motivating future careful treatment of this important process.  
\end{abstract}

\begin{keywords} 
planets and satellites: formation, gaseous planets  -- protoplanetary discs -- stars: formation \vspace{-2mm}
\end{keywords}


\section{Introduction}

Planets form in protoplanetary discs of dust and gas, growing and migrating within this disc during the first few Myr of a star's lifetime \citep[e.g.][]{Hai01}. Planet formation involves numerous processes that can govern early evolution, with the complicated interplay of these processes responsible for the final planet population \citep[for examples see reviews by][]{Baruteau14, Benz14, Drazkowska22}. {However, young stars preferentially occupy aggregates with an enhanced density of neighbouring stars with respect to the galactic field \citep[e.g.][]{Miller78,Larsen02,Lada03, Adamo17}.} In such environments, neighbouring stars may play a role in shaping the final planetary systems \citep[e.g.][]{Gil00, Brucalassi16, Winter20c}. In this work, we explore the role of irradiation from neighbouring OB stars in star forming regions on the growth and migration of massive planets. 

External photoevaporation is a process by which the heating by far ultraviolet (FUV) and extreme ultraviolet (EUV) photons from neighbouring OB stars heat the disc and drive thermal winds from the outer edge \citep{Johnstone98, Sto99}. This wind depletes the gas \citep{Ada04, Fac16, Haw18b}, and to some extent also the dust content \citep{Miotello12,Ans17, Sellek20}. Discs in several local star forming regions experience sufficient FUV fluxes to rapidly deplete discs \citep[e.g.][]{Ode94, Stolte04, Kim16,Gua16,  Win18b, Haworth21}. {The bright, massive and nearby discs occupying Taurus and Lupus are convenient to target for spatially resolved observations \citep[e.g.][]{HLTau_15}, but are known to occupy atypically low FUV environments for star formation.} Significant disc depletion occurs for approximately half of young stars in the solar neighbourhood \citep{Fat08, Winter20a}. 

Although external photoevaporation is a common process that influences the protoplanetary disc, it remains unclear to what degree it shapes the planets that form. This is particularly uncertain for planets that form close to their host star, where depletion {by external photoevaporation should be less vigorous} \citep[e.g.][]{Johnstone98}. Inner disc lifetimes appear to be shortened due to FUV irradiation in at least some regions \citep{ Stolte04,Fang12, Gua16}. At what separations and to what degree planet formation is influenced by externally driven winds remains an important open question. It is possible that the depletion of gas enhances dust-to-gas ratios \citep{Throop05, Haw18}, instigating the streaming instability \citep{Youdin05}. On the other hand, temperature variations may alter the efficiency of pebble accretion \citep{Ndugu18} and surface density gradients influenced by external photoevaporation can result in the rapid inwards migration of the largest dust grains \citep{Sellek20}. In this paper, we focus on the role of external photoevaporation for altering the growth and migration of planets that are accreting a massive gaseous envelope. Our models differ from the implementation in current population synthesis models \citep[e.g.][]{Emsenhuber21} due to our more detailed implementation of the FUV driven mass-loss and treatment of type II migration. We adopt theoretical mass-loss rates \citep{Haw18b} and winds that deplete the disc from the outer disc edge. 

Giant planet growth and migration has been studied extensively by numerous authors \citep[e.g.][see \citealt{Baruteau14} for a review]{Goldreich80, Veras04, Alexander09}. Indeed, internally driven photoevaporative winds are known to be capable of significantly altering planet migration outcomes \citep{Matsuyama03, Alexander12b,Coleman16,Jennings18,Kimmig20, Monsch21}. However, the role of the external wind on this process has yet to be explored; this is the focus of this work. To this end, we present a numerical model for the growth and migration of giant planets in externally irradiated discs in Section~\ref{sec:num_method}. We present the outcomes of our models in Section~\ref{sec:results} and draw conclusions in Section~\ref{sec:conclusions}.

\section{Numerical method}
\label{sec:num_method}
We apply a numerical method for the growth and migration of planets, which is simplified to concentrate on the role of external disc depletion on the accretion of gas onto a giant planet core {for a fixed internal disc wind profile}. In brief, we include prescriptions for type I and II migration, internal photoevaporation and viscous evolution, as well as the external depletion due to ambient FUV fields. We do not include the core accretion phase, choosing instead to inject ready-made `cores' at some time $\tau_\mathrm{form}$ \citep[as in][for example]{Veras04, Alexander09}. This is partly because the time-scales and physics of the core accretion stage are dependent on complex dust physics \citep[e.g.][]{Birnstiel12} and how this inter-plays with processes such as pebble accretion \citep[e.g.][]{Rafikov04, Lambrechts12, Drazkowska21}, dust traps \citep[e.g.][]{Barge95, Pinilla12}, streaming instability \citep[e.g.][]{Youdin05} and planetesimal accretion \citep{Ida04a,Alibert05,Coleman14}. The main aim of this work is to quantify the role of external photoevaporation on the massive planet population. {We therefore include only gas accretion physics}, encapsulating the uncertainties in the initial stages of formation by an assumed formation time-scale $\tau_\mathrm{form}$.

\subsection{Disc evolution}

We consider a gas disc that evolves under the evolution of viscous torques, internal and external photoevaporation and accretion onto a forming planet. The gas surface density $\Sigma_\mathrm{g}$ evolves according to the standard viscous diffusion equation \citep{Lynden-Bell74} with a viscosity $\nu$ expressed with the standard $\alpha$-parameter:
\begin{equation}
    \nu(r) = \alpha c_{\rm{s}}^2/\Omega_\mathrm{K}
\end{equation}at radius $r$, where $\Omega_\mathrm{K}$ is the (Keplerian) orbital frequency and $c_\mathrm{s}$ is the sound speed. The overall rate of change of the disc surface density is:
\begin{equation}
\label{eq:disc_evol}
   \partial_t {\Sigma_\mathrm{g}} = \frac 1 r \partial_r \left[  3 r^{1/2} \partial_r \left( \nu \Sigma_\mathrm{g} r^{1/2} \right) - \frac{2\Lambda \Sigma_\mathrm{g} r^{3/2}}{(GM_*)^{1/2}}\right] - \dot{Q}_\mathrm{p} - \dot{\Sigma}_{\rm{int}} -\dot{\Sigma}_{\rm{ext}}, 
\end{equation}where $\dot{Q}_\mathrm{p}$, $\dot{\Sigma}_{\rm{int}}$ and $\dot{\Sigma}_{\rm{ext}}$ are the surface density changes due to accretion onto the planet, internally driven winds and externally driven winds respectively. The second term on the RHS corresponds to the torque {exerted} by the planet, where $\Lambda$ is the rate of specific angular momentum transfer from the planet to the {disc}. We assume $\Lambda=0$ everywhere until a gap has been opened up (see Section~\ref{sec:migration}). 

We define $c_\mathrm{s}$ by adopting a temperature profile \citep{Hayashi81}:
\begin{equation}
    T = T_0 \left( \frac{r}{1\,\rm{au}}\right)^{-1/2} \left( \frac{L_*}{L_\odot}\right)^{1/4},
\end{equation}where $T_0 = 280$~K and in our fiducial model we have $L_* = 1\,L_\odot$. The isothermal sound speed is then related to the temperature via  $c_\mathrm{s} = \sqrt{k_\mathrm{B}T /m_\mathrm{H}}$, for $m_\mathrm{H}$ the mass of hydrogen. We then have $\nu\propto r$ for a constant $\alpha$ \citep[e.g.][]{Hartmann98}.

\subsubsection{Initial disc conditions and computation}

We generally consider an initial disc with a surface density profile:
\begin{equation}
\label{eq:Sigma0}
    \Sigma_\mathrm{g} = \Sigma_0 \left[1- \sqrt{\frac{r_\mathrm{in}}{r}
}\right] \left( \frac{r}{1\,\rm{au}} \right)^{-1} \exp \left[ -\frac{r}{R_\mathrm{s,0}} \right],
\end{equation}where $\Sigma_0 = 1700$~g~cm$^{-2}$ is the minimum mass solar nebula surface density at $1$~au \citep{Hayashi81}, $r_\mathrm{in} = 0.04$~au is the inner disc radius, $R_\mathrm{s,0} = 100$~au is the initial scale radius. {This power-law profile has the advantage of being simple, although it does not necessarily {capture} the true profile of physical discs \citep{Bitsch15a}.} Outside of the initial truncation radius $R_\mathrm{out,0}=200$~au we adopt a surface density floor in $\Sigma_\mathrm{g}$. This truncation radius is deliberately chosen to be large; models which include external FUV fields will quickly erode the disc down to smaller radii \citep[e.g.][]{Haw18b}, while the initial outer edge is unimportant for non-irradiated disc models. {The total initial disc mass with these parameters is $0.1 \, M_\odot$.}

For discs that are initially more compact the initial mass loss to the external wind may be lower. However, if the disc is subject to viscous torques we expect the photoevaporative outflow to balance the viscous mass-flux after approximately viscous time-scale \citep[e.g.][]{Winter20b, Hasegawa21}. {In addition, since the disc-planet torque operating on the planet scales steeply with distance from it (Section~\ref{sec:migration}), even when a disc is initially compact we expect our results to be weakly dependent on the choice of outer radius (except in inhibiting planet formation at $a_\mathrm{p,0}>R_\mathrm{out}$).} 

We solve the governing equation for the disc surface density (equation~\ref{eq:disc_evol}) over 1000 grid cells out to $r_\mathrm{out}=300$~au, equally spaced in $r^{1/2}$. We choose a time-step that is limited by the minimum of the time-scales to clear any cell of material via viscosity, {the photoevaporative winds} or the planet torque. In addition, we ensure that the planet mass and semi-major axis cannot change by more than 10~percent in a given time-step (although in practice the time-step is always set by the disc evolution). To avoid extremely short time-steps, we impose a minimum $r \Sigma_\mathrm{g}$ to be $10^{-8}$ times the maximum of $r\Sigma_\mathrm{g}$ at a given time, below which cells are assumed to have zero surface density. These choices do not appreciably influence the outcomes of our calculations. We perform the calculation until the total disc mass is $M_\mathrm{disc}<0.1 \,M_\mathrm{p}$ for $M_\mathrm{p}$ the mass of the planet (depletion of the disc), the planet semi-major axis $a_\mathrm{p}<0.15$~au (accretion of the planet onto the host star), or the time exceeds $10$~Myr.

\subsubsection{Planet accretion}

We include only the accretion of gas onto the planet, and define this accretion to change the surface density across a range $a_\mathrm{p}-R_\mathrm{H}< r< a_\mathrm{p}+R_\mathrm{H}$, where $a_\mathrm{p}$ is the semi-major axis of the planet and 
\begin{equation}
    R_\mathrm{H}= a_\mathrm{p} \left( \frac{M_\mathrm{p}}{3 M_*}\right)^{1/3}
\end{equation} is the Hill radius. The rate of change is such that:
\begin{equation}
  \int_0^\infty \dot{Q}_\mathrm{p} \cdot 2\pi r\,\mathrm{d} r = \dot{M}_\mathrm{p},
 \end{equation}where $\dot{M}_\mathrm{p}$ is the accretion rate of the planet (see Section~\ref{sec:accretion}). The form of $\dot{Q}_\mathrm{p}$ is discussed in Section~\ref{sec:accretion}, and is chosen to be mass conservative.

\subsubsection{Internal wind and viscosity}
\label{sec:int_wind}

\begin{figure}
    \centering
    \includegraphics[width=\columnwidth]{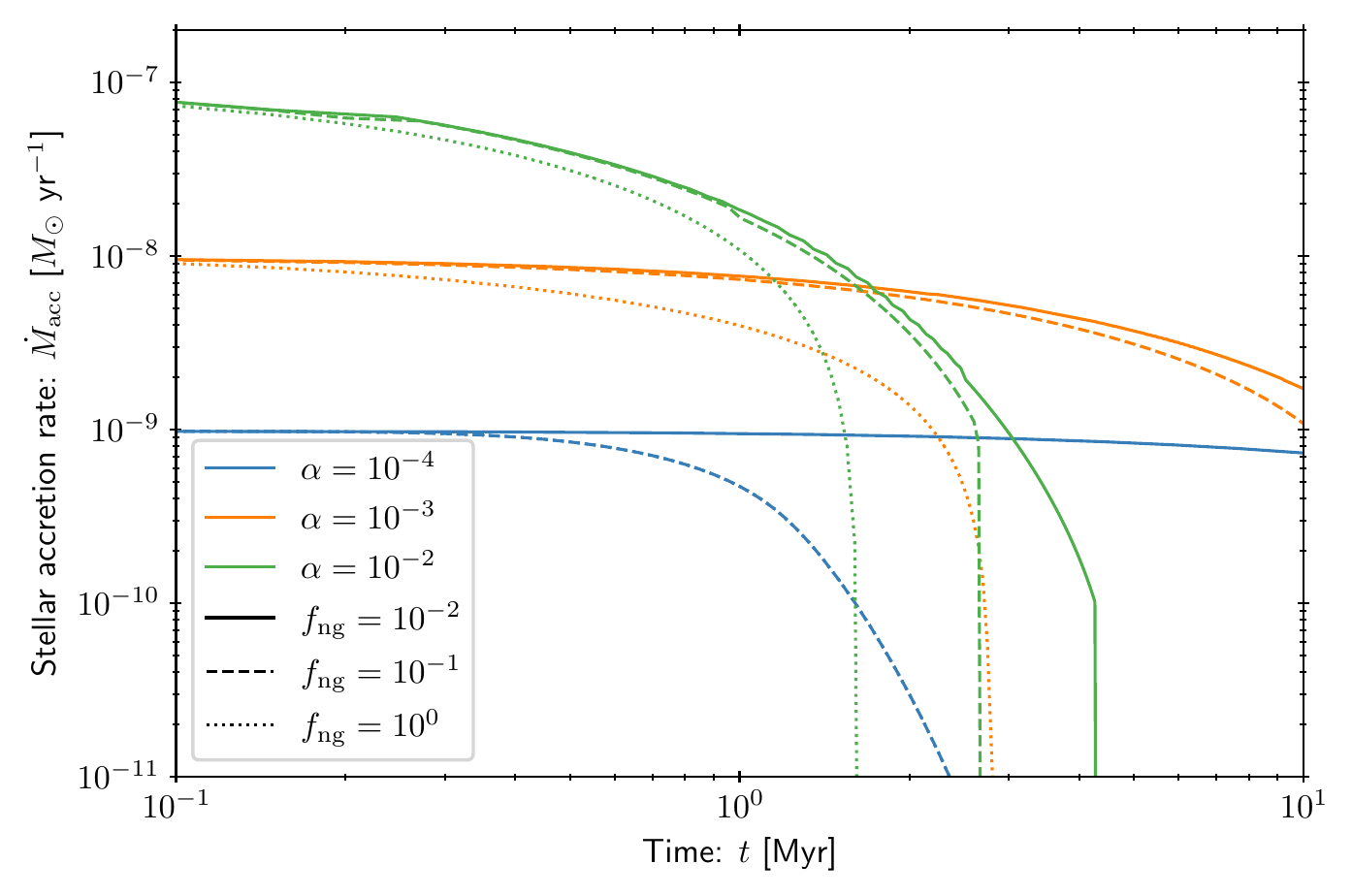}
    \caption{Accretion rate evolution in an isolated $\alpha$-disc model with no external FUV flux. The internal driven disc wind is assumed to drive a mass loss rate $\dot{M}_\mathrm{int} = 10^{-8}$~$M_\odot$~yr$^{-1}$ when a gap has been opened, while this is supressed by a factor $f_\mathrm{ng}$ prior to gap opening. We vary $\alpha$ and $f_\mathrm{ng}$. {The line for $\alpha=10^{-4}$ and $f_\mathrm{ng}=0.1$, $1$ are the same, since the initial accretion rate is below $0.1 \cdot 10^{-8} \,M_\odot$~yr$^{-1}$.}}
    \label{fig:Mdot_acc}
\end{figure}

The internally driven wind can generally be launched either thermally, by photons that heat the gas \citep[see][for a review]{Ercolano17}, by magnetohydrodynamic pressure gradients \citep[e.g.][]{Suzuki09}, or some combination \citep{Ballabio20, Rodenkirch20}. {If winds are photoevaporative, X-ray photons are expected to penetrate to large column densities, driving vigorous flows that dominate mass-loss over EUV driven winds \citep[e.g.][]{Pascucci14}. Whether X-ray photoevaporation is suppressed depends on the column depth experienced by photons. Surface densities $\sim 10^{20}$~cm$^{-2}$ are required to suppress the flow \citep{Owen12}, otherwise mass-loss in the X-ray driven wind should be $\sim 10^{-8}\,M_\odot$~yr$^{-1}$. This is probably the case if X-ray heating is dominated by photons of energy $\sim 0.3{-}0.4$~keV, as supported by the findings of \citet{Owen10}. However, internal processes such as dust grain evolution in the inner disc may also play an important role on which photons drive photoevaporative winds at each stage in the disc lifetime \citep[see][for a discussion of late stage {photoevaporation}]{Nakatani21}. }

{Several authors have performed detailed studies on how planet migration varies with the internally driven wind profile \citep[e.g.][]{Alexander12b, Ercolano15, Jennings18, Monsch21b}. These studies frequently find that if the (range of) initial conditions is correct, winds can halt migration close to the wind-launching radius \citep{Matsuyama03}, resulting in a dearth of massive planets at this location.} {Rather than explore the possible internal wind profiles \citep[see][for example]{Jennings18}, here we fix the internal mass-loss rate profile. We vary only the external FUV flux, allowing us to isolate the role of the externally driven winds. In doing so, we also demonstrate that the known range of disc environments can produce diverse planet formation scenarios, without the need for appealing to diverse internal properties.} Our internal wind model is similar to that of \citet{Cla01}. We assume that the base density of an ionised, internally driven photoevaporative flow should approximately follow \citep[e.g.][]{Hol94}:
\begin{equation}
    n_0 \propto r^{-5/2},
\end{equation}and the corresponding surface density rate of change is:
\begin{equation}
    \dot{\Sigma}_\mathrm{int} \approx n_0 c_\mathrm{s,w},
\end{equation}where $c_{\rm{s,w}} \approx 10$~km~s$^{-1}$ is the sound speed in the (ionised) wind. 

The wind is launched outside of a radius $R_\mathrm{launch} \approx 0.2 R_\mathrm{g}$, where
\begin{equation}
    R_\mathrm{g} = \frac{G M_*}{c_{\rm{s,w}}^2} 
\end{equation}{is the gravitational radius. The radius $  R_\mathrm{g}$ is the radius beyond which the thermal energy in the gas exceeds the gravitational binding energy. The radius $R_\mathrm{launch} $ is somewhat smaller than this due to the contribution of hydrodynamic effects \citep{Liffman03}.} In practice, when the inner edge of the disc $R_\mathrm{in}$ exceeds $0.2 R_\mathrm{g}$, we adopt $R_\mathrm{in}$ as the launching radius instead. 

We adopt a fixed mass-loss rate throughout the later stages of disc evolution, since the mass-loss rate should be relatively weakly dependent on the inner edge radius \citep{Owen10}. Once an inner gap has opened up, we assume that X-ray photons dominate the mass-loss rates, yielding a total mass-loss \citep[e.g.][]{Owen10}:
\begin{equation}
\dot{M}_\mathrm{int, gap} = \int_{R_\mathrm{launch}}^\infty \dot{\Sigma}_{\rm{int}} \cdot 2\pi r \,\mathrm{d}r \approx  10^{-8} \left(\frac{L_\mathrm{X}}{10^{30} \,\rm{erg~s}^{-1}}\right) \,M_\odot \rm{yr}^{-1},
\end{equation} where $R_\mathrm{launch}$ is the launching radius and {we adopt $L_\mathrm{X}=10^{30}$~erg~s$^{-1}$, which is the X-ray luminosity of the host star.}

{However, a viscous disc model with $\dot{M}_\mathrm{int, gap} \sim 10^{-8}\, M_\odot$~yr$^{-1}$ is inconsistent with observed disc properties. To produce the observed disc life-times of a few Myr \citep[e.g.][]{Hai01} and accretion rates commonly found down to $\sim 10^{-11}{-}10^{-10}\, M_\odot$~yr for solar mass stars, this mass-loss rate must be initially suppressed. It is not clear how such a suppression might operate, although accretion streams and dust evolution may alter photoevaporation rates \citep[][]{Owen12, Nakatani21}. {Here we introduce a suppression factor $f_\mathrm{ng}$ in the total mass-loss rate prior to the gap opening, such that:}
\begin{equation}
    \dot{M}_\mathrm{int} = \begin{cases}
   f_\mathrm{ng} \cdot \dot{M}_\mathrm{int, gap}  \qquad & \dot{M}_\mathrm{acc} >  f_\mathrm{ng} \cdot \dot{M}_\mathrm{int, gap}    \\
   \dot{M}_\mathrm{int, gap}  \qquad & \rm{otherwise}
    \end{cases}.
\end{equation}When $\dot{M}_\mathrm{int} > \dot{M}_\mathrm{acc}$ we increase the mass-loss rate to $\dot{M}_\mathrm{int, gap}$, emulating the transition from an optically thick inner disc to an optically thin one.  }

In Figure~\ref{fig:Mdot_acc} we show the accretion rates over the disc lifetime for various values of $\alpha$ and $f_\mathrm{ng}$. {Regardless of internal photoevaporation, accretion rates decrease with time as the disc viscously evolves. Once the accretion rate drops sufficiently, the inner disc is rapidly cleared due to the internal wind. This phase of rapid clearing is seen statistically as a dearth of accretion rates below some threshold \citep{Manara17}, thus the maximum rate of internal photoevaporation for a young disc can be empirically estimated. }

We here generally adopt $\alpha=3\cdot 10^{-3}$ and $f_\mathrm{ng}=0.1$, which reproduces an inner disc lifetime that is similar to that of local star forming regions \citep{Hai01}. The observed distribution of accretion rates, including many solar mass stars with accretion rates down to $\sim 10^{-10} \, M_\odot$~yr$^{-1}$, are also broadly consistent with these choices \citep[e.g.][]{Manara16, Manara17}. {We discuss our choices in Appendix~\ref{app:int_wind_discuss}, showing that varying $f_\mathrm{ng}$ has little influence on the outcomes of our fiducial model parameters in the limit of small external flux ($F_\mathrm{FUV}=0\, G_0$). This is not the case for all initial conditions and parameter choices; many authors have demonstrated that under certain conditions the internal wind can halt planet migration \citep{Matsuyama03, Ercolano15}. Thus, for certain ranges of initial conditions, internal photoevaporation produces a dearth of planets at the wind launching radius \citep[e.g.][]{Jennings18, Monsch21b}. However, the goal of this work is not a synthesis study, rather demonstrating that the range of external FUV fluxes that young discs experience can produce a diversity without appealing to large variations between isolated star-disc systems. }

Finally, we note that large $\alpha$ values appear inconsistent with the geometry of dust substructure in at least some protoplanetary discs \citep{Pinte16}. The disc evolution may instead be dictated by magnetohydrodynamic (MHD) winds rather than thermal ones \citep{Tabone21}. Speculatively, we would expect a similar mass flux through the disc, which must be required to sustain the observed accretion rates. The main influence of MHD winds might be in the gap opening criteria (see Section~\ref{sec:migration}) and the nature of disc dispersal. How this may alter our findings will be explored in future work.


\subsubsection{External wind}

The externally driven wind represents the novel component of our models. In dense stellar aggregates, both EUV and FUV photons from massive neighbouring stars can launch winds that yield rapid mass-loss rates of up to $\dot{M}_\mathrm{ext} \sim 10^{-6}\,M_\odot$~yr$^{-1}$ from the outer disc edge \citep{Ode94, Johnstone98, Ada04, Fac16}. EUV photons dominate the mass-loss rate in the limit of a thin photodissociation region (PDR), which can occur very close to, or far from, an ionising source \citep{Win18b}. However, the long-term disc evolution should generally be dominated by the FUV photons. This is because the EUV driven wind $\dot{M}_\mathrm{ext, EUV}$ {scales with} $R_\mathrm{out}^{3/2}$, for outer disc radius $R_\mathrm{out}$. {Meanwhile, the mass-loss from the FUV driven disc wind scales approximately linearly: $\dot{M}_\mathrm{ext, FUV} \propto R_\mathrm{out}$}. Thus, as the outer part of the disc is eroded the disc is eventually compact enough such that FUV-driven mass-loss dominates. 

In order to model the influence of external photoevaporation on the disc evolution, the surface density is evolved in a similar way to previous studies \citep[e.g.][]{Cla07, Ande13, Win18b}. In brief, this involves identifying the cell at the outer edge and advancing or retreating this edge depending on whether there is net accumulation (due to viscous expansion) or depletion (due to external mass-loss). We adopt the FUV mass-loss rates in the `FRIED' grid \citep{Haw18b}, which are computed by coupling the thermal structure in the PDR to thermal wind solutions \citep[e.g.][]{Fac16}. For a given stellar host mass, we interpolate these rates over the instantaneous disc mass, radius and FUV flux (measured in the \citeauthor{Hab68} unit of $1\,G_0\equiv 1.6\times 10^{-3}$~erg~cm$^{-2}$~s$^{-1}$). Strictly, the important quantity in determining the mass-loss rate is the outer surface density. However, we use the total disc mass and outer radius as a proxy since the steady state surface density profile of our disc scales with $r^{-1}$, which is the same as that of \citet{Haw18b}. Using the total mass enables a greater stability, since we are not sensitive to the fluctuations in the surface density at the grid cell that is defined as the outer radius.

\subsection{Planet growth}
\label{sec:accretion}

{We do not treat the initial planet growth from the solid disc content, which is a complex process involving numerous physical processes operating across many orders of magnitude in scale.} However, we focus here on the final stages of mass accumulation, where the planet undergoes runaway accretion of gas. \citet{Ida04a} estimate a threshold mass above which atmospheric pressure no longer supports {the} gas envelope against the planet gravity \citep{Ikoma00}:
\begin{equation}
\label{eq:Mcrit}
    M_\mathrm{crit} \sim 10 \left( \frac{\dot{M}_\mathrm{core}}{10^{-6} \, M_\oplus~\rm{yr}^{-1}} \right)^{0.25} \, M_\oplus.
\end{equation}This threshold is dependent on the accretion rate in order that the energy of the accreted mass does not prevent hydrostatic collapse \citep{Bodenheimer86, Pollack96}. Since the ability for a gaseous envelope to collapse is dependent on its ability to irradiate thermal energy, $ M_\mathrm{crit}$ should also depend on the envelope opacity. This dependence is neglected in equation~\ref{eq:Mcrit} because of the inherent uncertainties, however $M_\mathrm{crit}\sim 5{-}20$~$M_\oplus$ is a reasonable canonical estimate \citep{Ikoma00}. In this work, we do not implement a critical mass threshold, since this this is necessarily dependent on the range of processes that govern solid growth -- such as pebble accretion, dust traps and the streaming instability \citep[e.g.][]{Lambrechts14,Drakowska14,Drazkowska16, Voelkel21}.  Instead, we will initiate planets with $M_\mathrm{p}=10 \,M_\oplus$, approximately the lowest mass at which runaway growth should be instigated, at some time $\tau_\mathrm{form}$ during the disc lifetime. 


Once formed, we follow \citet{Ida04a, Ida04b} in adopting a planet growth rate:
\begin{equation}
    \dot{M}_\mathrm{p} = \min \left\{\dot{M}_\mathrm{KH}, \epsilon \dot{M}_\mathrm{flux}\right\},
\end{equation}where $\dot{M}_\mathrm{KH}$ is the Kelvin-Helmholtz contraction timescale of the envelope:
\begin{equation}
   \dot{M}_\mathrm{KH} =   \dot{M}_\mathrm{KH,0} \left(\frac{M_\mathrm{p}}{10\, M_\oplus}\right)^{k}
\end{equation}with $k=4$ and $\dot{M}_\mathrm{KH,0} = 10^{-5}  M_\oplus$~yr$^{-1}$. The growth rate due to the viscous mass flux into the annulus defined by the planet's Hill radius $R_\mathrm{H}$ is:
\begin{equation}
    \dot{M}_\mathrm{flux} =\dot{M}_\mathrm{flux,out}+ \dot{M}_\mathrm{flux,in}
\end{equation}where 
\begin{equation}
    \dot{M}_\mathrm{flux,out} = \max \{F_\mathrm{visc}(a_\mathrm{p}+C_\mathrm{H}R_\mathrm{H}),0\},
\end{equation}
\begin{equation}
    \dot{M}_\mathrm{flux,in} = \max \{-F_\mathrm{visc}(a_\mathrm{p}-C_\mathrm{H}R_\mathrm{H}),0\},
\end{equation}
and
\begin{equation}
\label{eq:flux}
    F_\mathrm{visc}(r) = 3\pi\nu \Sigma_\mathrm{g}+ 6\pi r \partial_r \left(\Sigma_\mathrm{g} \nu \right).
\end{equation} The pre-factor $\epsilon$ we adopt is not identical to that of \citet{Ida04a, Ida04b} who use an approximate expression for the efficiency of growth but do not consider the flux of material across a gap.  We are explicitly interested in the quantity of material moving across the gap, for which we follow the parametrisation by \citet[][see also \citealt{Veras04}]{Alexander09}:
\begin{equation}
    \frac{\epsilon}{\epsilon_\mathrm{max}} = 1.67 \left(\frac{M_\mathrm{p}}{1\, M_\mathrm{J}}\right)^{1/3}\exp \left(- \frac{M_\mathrm{p}}{1.5 \, M_\mathrm{J}}\right) +0.04, 
\end{equation}where $\epsilon_\mathrm{max}=0.5$ and this formula is chosen to approximate the numerical results of \citet{Lubow99} and \citet{Dangelo02}. 

{The factor $C_\mathrm{H}$ sets the width of the feeding zone, for which we will generally adopt $C_\mathrm{H}=1$. Previous studies of growth and type II migration have often taken the viscous mass flux further from the planet. For example, \citet{Alexander09} adopt the steady state viscous flux at $r=3a_\mathrm{p}$ to estimate the accretion rate onto the planet. \citet{Coleman14} estimate that adopting $C_\mathrm{H}=10$ reproduces the results of higher dimensional numerical simulations. However, both of these works assume the steady state viscous flux $ F_\mathrm{visc} = 3\pi\nu \Sigma_\mathrm{g}$, which is spatially constant at large radii for an isolated disc after a viscous time-scale. In this case, we cannot apply a similar prescription because the outer edge of the disc is depleted by the wind. Adopting the viscous flux at large radii can therefore prematurely shut off planet growth. For this reason, we adopt the smaller feeding zone radius with $C_\mathrm{H}=1$, but without the assumption of steady state (i.e. with flux defined by equation~\ref{eq:flux}). For a disc not subject to external depletion, this yields a similar accretion rate as the steady state estimate while the viscosity in the disc is sufficient to overcome the torque from the planet and replenish material at the edge of the gap. Since the disc outside the gap can be externally depleted, it is also necessary to consider the mass flow from both inside and outside the gap.  }

 {To treat the flow of mass across the gap, we adopt an approach that is similar to that of \citet{Alexander09}, but generalised to allow a flow in both directions. In practice, this is achieved by computing the quantity of material flowing inwards and outwards, but not being accreted onto the planet ($\dot{M}_\mathrm{flux,out}/(1+\epsilon)$ and $\dot{M}_\mathrm{flux,in}/(1+\epsilon)$ respectively). This is balanced with the total mass removal from the relevant edge to establish whether there is a net mass increase or decrease. In the case of net mass loss, the relevant amount of mass is removed from the cell(s) closest to the planet -- i.e. progressively from the gap edge until the requisite mass is reached. In the case of net mass gain, the extra mass is placed at the edge of the feeding zone on the appropriate side of the gap. This scheme conserves the total mass of the system. We also adjust the semi-major axis {of the planet} such that the angular momentum of the material that is accreted goes into the planet orbit. }


\subsection{Migration}
\label{sec:migration}
{For type I migration, before a gap has opened in the disc, we use the conventional formula for the rate based on the results of \citet{Paardekooper11}. The total torque $\Gamma_\mathrm{I}$ is the sum of the Lindblad torque, horseshoe drag torques, and corotation torques, as given by equations 3-7 in \citet{Paardekooper11}. We here adopt a locally isothermal, optically thin disc, for which the vorticity torques dominate. Adopting more complex dependence on the opacity with temperature \citep[e.g.][]{Henning96} can yield discontinuities in the torque and diverse migration behaviour \citep[e.g.][]{Coleman14}. However, in this work we do not wish to explore this complex behaviour and the dependence on opacity. It is therefore beneficial to maintain a simple form to isolate the influence of external disc depletion. The overall magnitude of the torque is:}
\begin{equation}
   | \Gamma_{\rm{I}} | \sim  \Gamma_{\rm{I},0} /\gamma,
\end{equation}where $\gamma$ is the adiabatic exponent and 
\begin{equation}
    \Gamma_{\rm{I},0} = q^2 \left( \frac{a_\mathrm{p}}{H_\mathrm{p}} \right)^2 \Sigma_\mathrm{p} \Omega_\mathrm{p}^2 a_\mathrm{p}^4
\end{equation}{and the subscript `p' represents the values at the location of the planet. This torque results in the rate of change of semi-major axis:}
\begin{equation}
    \dot{a}_\mathrm{p,I} = 2 \left( \frac{a_\mathrm{p}}{G M_*}\right)^{1/2} \frac{\Gamma_\mathrm{I}}{M_\mathrm{p}}.
\end{equation} 


{Population studies have found that this prescription for type I migration results in rapid accretion of low mass planets onto the central star. Thus factors $10^{-3}{-}0.1$ are applied in synthesis studies to mitigate this problem, while the origin of this apparent discrepancy remains an open question  \citep[see discussion by][]{Benz14}. Here we adopt a slow-down factor of $10^{-2}$. }

In order to begin opening a gap in the disc to transition to type II migration, the tidal torque of the planet must exceed the viscous stress and local thermal pressure. We adopt the gap opening criteria of \citet[][see also \citealt{Baruteau14}]{Crida06}:
\begin{equation}
\label{eq:gap_open}
    \frac{h_\mathrm{p}}{q^{1/3}} + \frac{50\alpha h_\mathrm{p}^2}{q} < 1,
\end{equation} where $h_\mathrm{p}= H_\mathrm{p}/a_\mathrm{p}$ is the aspect ratio of the disc at the location of the planet, and $q\equiv M_\mathrm{p}/M_*$ is the mass ratio of the planet to the host star. If equation~\ref{eq:gap_open} is satisfied or $\Sigma_{\rm{p}}$ {reaches the surface density floor}, then we assume a gap has been opened in the disc. The reason for the low surface density condition is that in some instances material may be cleared by the internal or external wind, rather than the rate at which viscosity is diffusing material into the gap. 

For the type II migration rate, we follow the prescription of \citet[][see also \citealt{Lin86}]{Alexander09}. Following \citet{Trilling98}, the specific angular momentum transfer to be substituted into equation~\ref{eq:disc_evol} is:
\begin{equation}
    \Lambda = \begin{cases} -\frac{q^2 G M_*}{2r} \left(\frac{r}{\Delta_\mathrm{p}} \right)^4 \qquad \qquad & r< a_\mathrm{p} \\
    \frac{q^2 G M_*}{2r} \left(\frac{r}{\Delta_\mathrm{p}} \right)^4 \qquad \qquad &  r \geq a_\mathrm{p}
    \end{cases},
\end{equation}where 
\begin{equation}
    \Delta_\mathrm{p} = \max \{ H, |r-a_\mathrm{p}| \}.
\end{equation}As in \citet{Alexander09}, when the gap is narrow it is necessary to place a limit on $|\Lambda|$ such that the torque time-step does not become prohibitively small and the resolution required for accurate computation becomes high. As for \citet{Alexander09}, we enforce an upper limit $|\Lambda|\leq 0.1 rH \Omega^2$. Such a limit is computationally necessary, but also reflects the  finite physical width over which the resonances operate; on small physical scales thermal effects can nullify their influence \citep{Lin86}. The migration of the planet then proceeds at a rate:
\begin{equation}
   \dot{a}_{\mathrm{p,II}} = - \left(\frac{a_\mathrm{p}}{G M_*}\right)^{1/2} \frac{4\pi}{M_\mathrm{p}} \int r\Lambda \Sigma \,\mathrm{d} r.
\end{equation}This concludes the prescriptions implemented in our planet-disc evolution model.

\subsection{Parameter choices}

Given the large number of parameters defining our model, it is not possible or helpful to perform a full parameter study. Instead we must choose some fiducial values, and vary some key parameters to test the dependence of our outcomes on these choices. Our choices are summarised in Table~\ref{tab:params}. 

\begin{table}
    \centering
    \begin{tabular}{c c c }
    \hline
         Symbol & Meaning  & Fiducial value \\
         \hline
        $\alpha$ & Viscosity constant & $3\cdot 10^{-3}$  \\
        $\tau_\mathrm{form}$ & Time of planet injection & 1~Myr  \\
        $M_*$ & Host star mass  & $1 \, M_\odot$  \\
        $M_{\mathrm{p},0}$ & Initial planet mass & $10 \, M_\oplus $ \\
        $a_{\mathrm{p},0}$ & Initial planet semi-major axis & $20  $~au \\
        $\dot{M}_{\mathrm{int}}$ & Internal wind mass-loss rate & $10^{-8} \, M_\odot $~yr$^{-1}$ \\
        $f_{\mathrm{ng}}$ & Early wind suppression factor  & $0.1$ \\
        $C_{\mathrm{I}}$ & Type I suppression factor & $10^{-2} $ \\
        $\Sigma_0$ & Surface density at $1$~au & $1700$~g~cm$^{-2}$ \\
        $T_\mathrm{0}$ & Disc temperature at $1$~au & $280$~K \\
        $r_\mathrm{in}$ & Inner grid radius & $0.04$~au \\
        $R_\mathrm{out,0}$ & Disc truncation radius & $200$~au \\
        $R_\mathrm{s,0}$ & Disc scale radius & $100$~au \\
        $\dot{M}_\mathrm{KH,0}$ &{KH contraction rate at $10\,M_\oplus$} & $10^{-5} \,M_\oplus$~yr$^{-1}$ \\
        $k$ & Index for KH contraction &  $4$ \\
        \hline
    \end{tabular}
    \caption{Fiducial parameters adopted in the planet evolution model.}
    \label{tab:params}
\end{table}

\section{Results and discussion}
\label{sec:results}
\subsection{Evolution of the planet core initially at 10~au}
\begin{figure}
    \centering
    \includegraphics[width=\columnwidth]{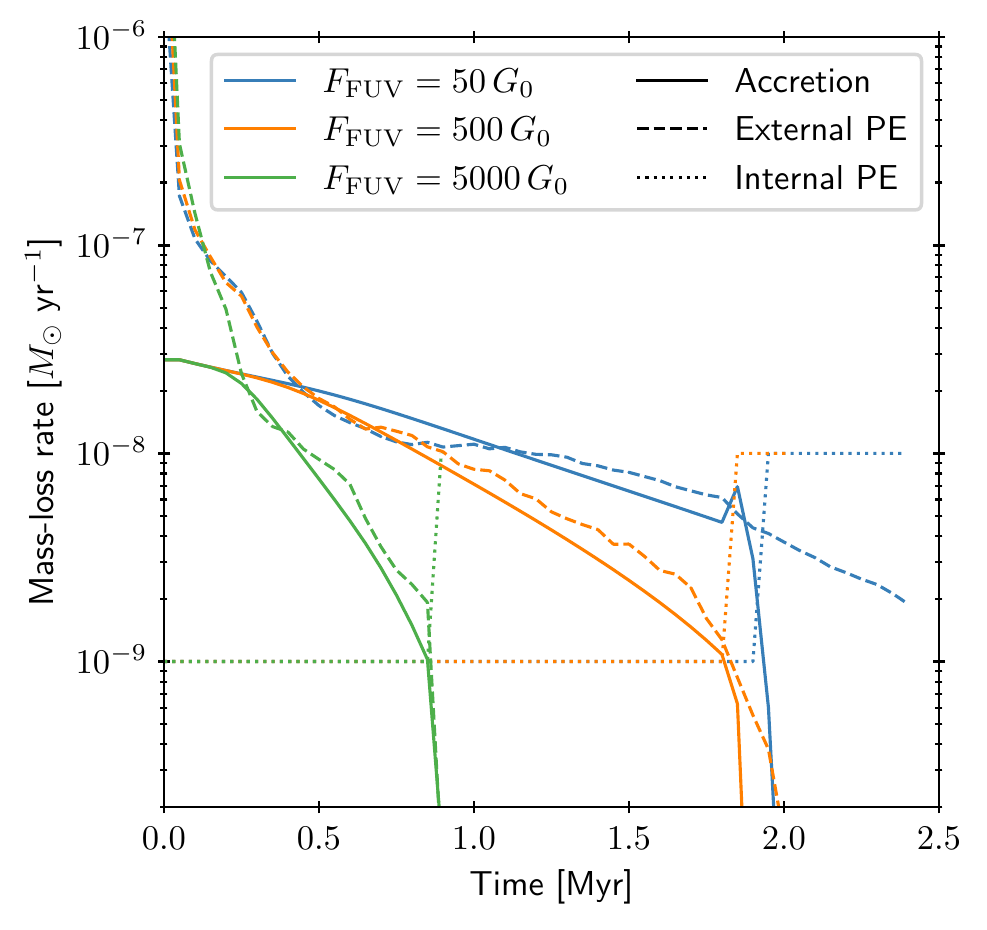}
    \caption{{The {totel} mass-loss rates for the fiducial disc model with parameters described in Table~\ref{tab:params}, with varying external FUV flux $F_\mathrm{FUV}$. We show the disc mass-loss rates due to accretion {onto the host star} (solid lines), external photoevaporation (dashed lines) and internal photoevaporation (dotted lines). The outcomes for $F_\mathrm{FUV} = 50$, $500$ and $5000 \, G_0$ are shown in blue, yellow and green respectively.}}
    \label{fig:mdots_20}
\end{figure}

\begin{figure*}
    \centering
    \subfloat[Planet evolution\label{fig:aM_evol}]{\includegraphics[width=0.87\columnwidth]{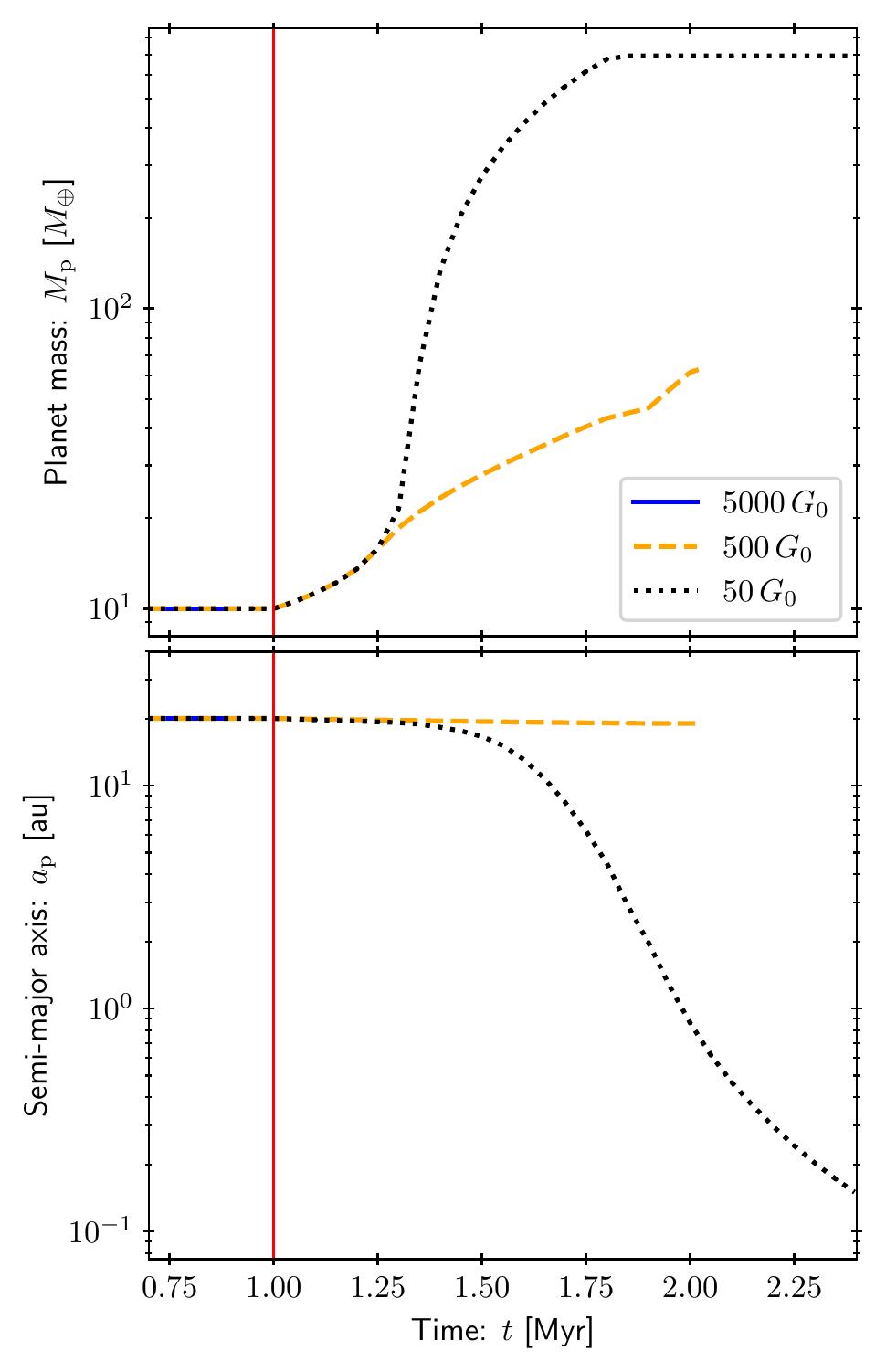}}
    \subfloat[Surface density evolution\label{fig:sd_evol}]{\includegraphics[width=1.13\columnwidth]{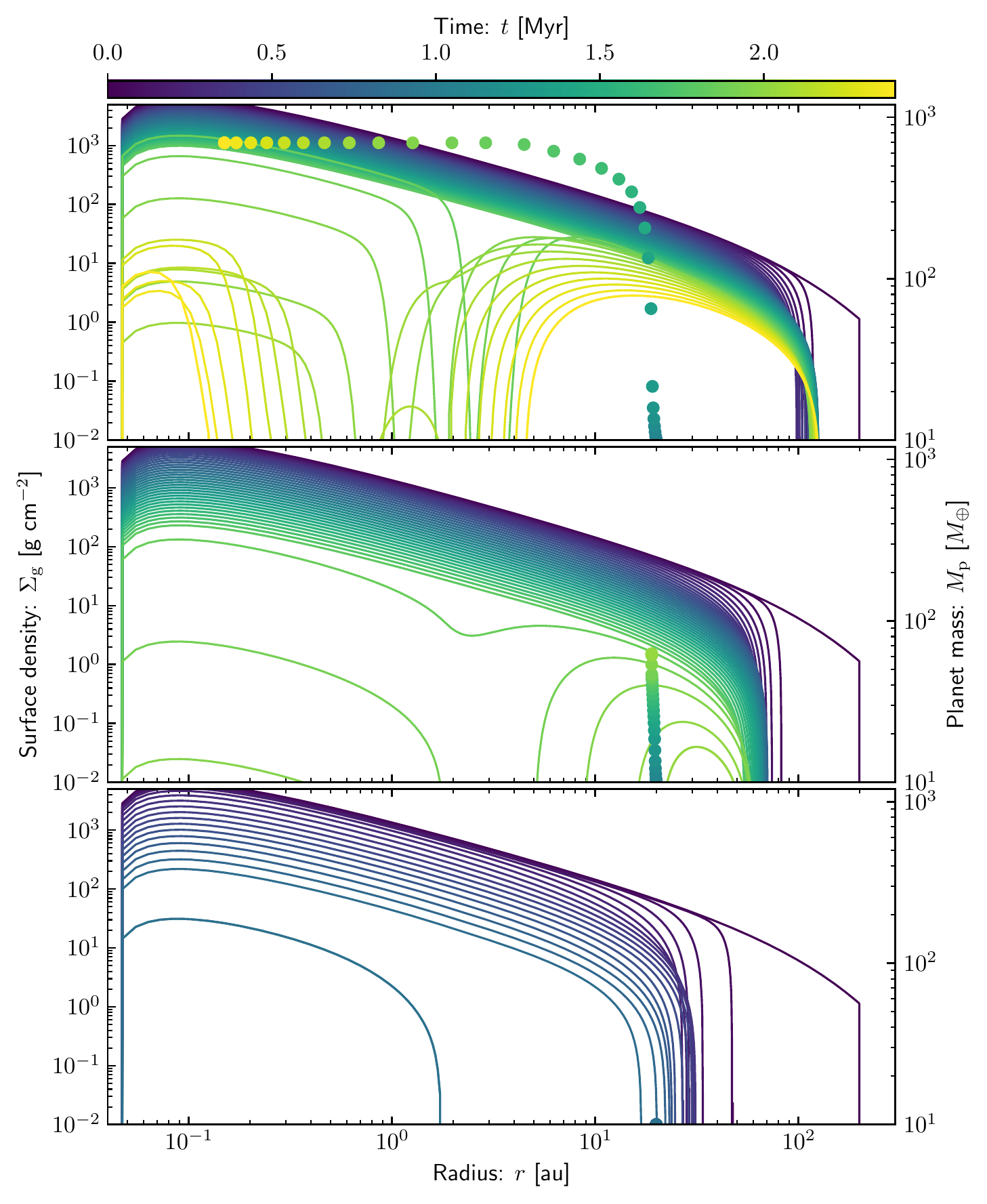}}
    \caption{Planet and disc surface density evolution for our fiducial disc and planet model under the influence of different FUV fluxes (as in Figure~\ref{fig:mdots_20}). In Figure \ref{fig:aM_evol} we show the evolution of the planet mass (top) and semi-major axis (bottom) for an initial mass $M_\mathrm{p,0} = 10\,M_\oplus$ and semi-major axis $a_\mathrm{p,0} = 20$~au. The red vertical line indicates the time at which the planet is injected, $\tau_\mathrm{form}=1$~Myr. Dotted, dashed and solid lines represent FUV fluxes of $F_\mathrm{FUV} = 50 , 500$ and~$5000$~$G_0$ respectively. Our fiducial disc has $\alpha=3\cdot 10^{-3}$ and $f_\mathrm{ng} = 0.1$ --  see text for details. In Figure~\ref{fig:sd_evol} we show the corresponding surface density evolution at the same three different FUV fluxes $F_\mathrm{FUV}=50$, $500$ and $5000$~$G_0$ (top, middle and bottom respectively). The lines represent the surface density at radius $r$, while the circles represent the planet mass and semi-major axis ($a_\mathrm{p}=r$). Both are coloured by the time in the simulation shown by the colour bar (top), and are each spaced by $0.05$~Myr. } 
\end{figure*}

We begin by considering the outcome of our fiducial model at three different FUV fluxes: $F_\mathrm{FUV} = 50$, $500$ and~$5000$~$G_0$. {The mass-loss rates due to accretion, internal photoevaporation and external photoevaporation are shown in Figure~\ref{fig:mdots_20}. The externally driven mass-loss rate drops rapidly because the outer disc is initially easily unbound by the photoevaporative wind, leading to high loss rates. Eventually, the mass-loss rate in the wind is balanced by the viscous expansion, such that $\dot{M}_\mathrm{acc} \approx \dot{M}_\mathrm{ext}$. As discussed in Section~\ref{sec:int_wind}, mass-loss rate in the internal wind is held constant at $\dot{M}_\mathrm{int}= 10^{-9} \, M_\odot$~yr$^{-1}$ until the accretion rate is $\dot{M}_\mathrm{acc} < 10^{-9} \, M_\odot$~yr$^{-1}$. Subsequently, $\dot{M}_\mathrm{int}= 10^{-8} \, M_\odot$~yr$^{-1}$ for the remainder of the disc lifetime. In the case of the low flux, $F_\mathrm{FUV}=50\,G_0$ simulation, the accretion rate is strongly suppressed by the massive planet at late times. }

The growth and migration of the planet is shown in Figure~\ref{fig:aM_evol}, while this can be compared to the disc evolution in Figure~\ref{fig:sd_evol}. We find that the growth and migration behaviour strongly depends on $F_\mathrm{FUV}$. For low $F_\mathrm{FUV}=50\,G_0$, the planet grows rapidly until it opens a gap in the disc, at which point it undergoes type II inward migration. Growth is then slower due to the balance of the planet torque and viscous redistribution of material into the gap.

By contrast, for intermediate $F_\mathrm{FUV} = 500\,G_0$, we find that the planet does not undergo any substantial type II migration due to the low planet mass and rapid disc dispersal. The planet can undergo initial growth, but the disc is depleted before any appreciable inward migration. In principle, even outward migration in the type II stage is possible \citep[][]{Veras04}. 

{At high FUV fluxes ($F_\mathrm{FUV}\sim 5000\,G_0$), the disc is dispersed before the planet core is even injected into the system. In this limit, we expect that external photoevaporation should strongly influence the earlier processes of planetesimal formation that we do not treat in this work (see discussion in Section~\ref{sec:caveats}). However, we can conclude that in this high $F_\mathrm{FUV}$ limit giant planet growth and migration should be strongly influenced by the external depletion. }

We now discuss how the role of the external FUV flux varies with varying initial conditions.

\subsection{Varying initial planet semi-major axis}

\begin{figure*}
    \centering
    \includegraphics[width=\textwidth]{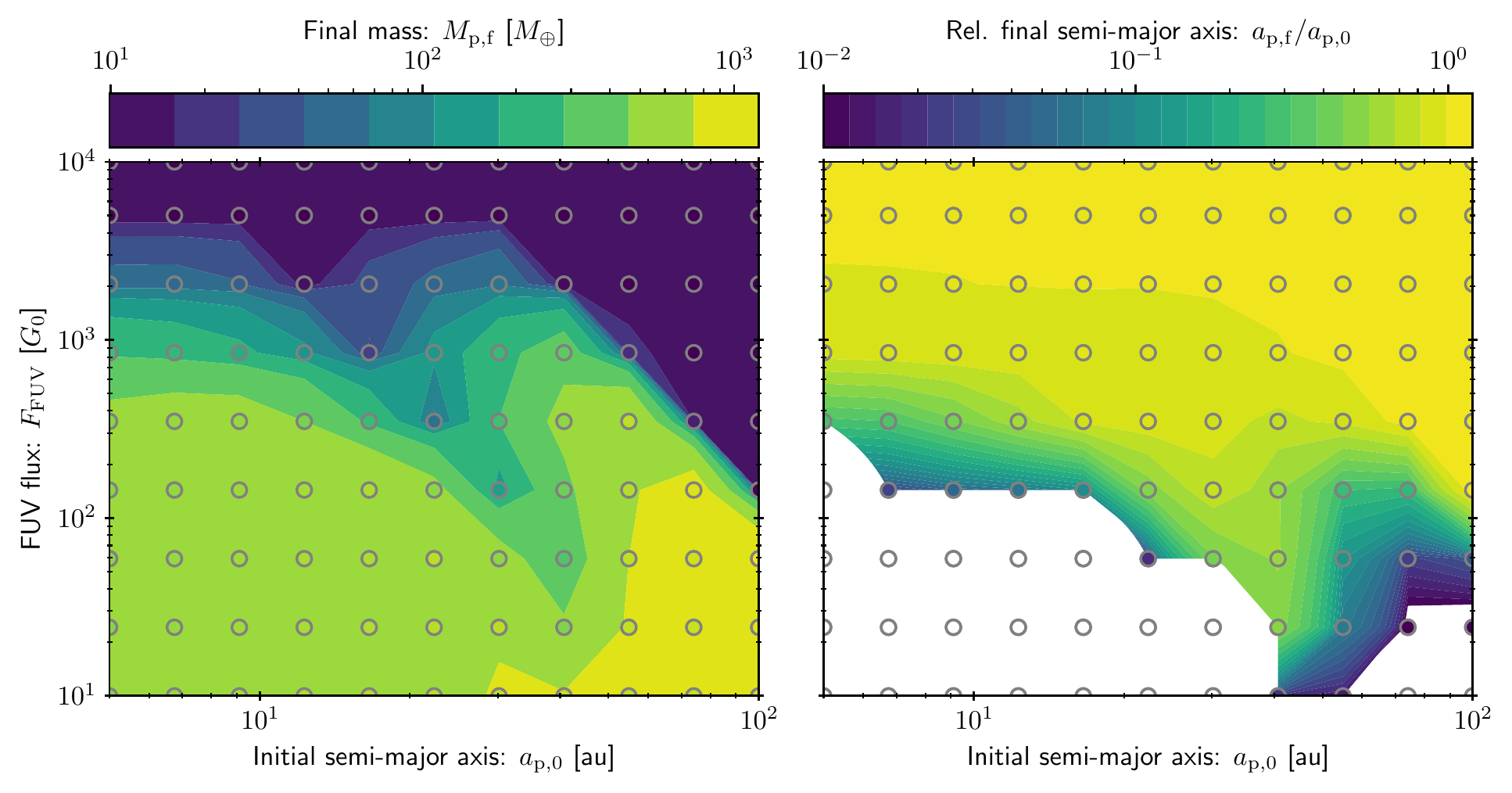}
    \caption{Parameter space exploration for the final mass ($M_\mathrm{p,f}$, left) and semi-major axis ($a_\mathrm{p,f}$, right) of a planet in the plane of initial planet semi-major axis $a_\mathrm{p,0}$ and FUV flux $F_\mathrm{FUV}$. All other parameters are fixed as described in Table~\ref{tab:params}. The final semi-major axis is shown as a fraction of the initial value: $a_\mathrm{p,f}/a_\mathrm{p,0}$. Grey circles show the grid points for which we calculate the planet evolution. {White regions are where the planet reaches a semi-major axis $a_\mathrm{p}<0.15$~au before the end of the disc lifetime.}}
    \label{fig:outcome_exp}
\end{figure*}

\begin{figure}
    \centering
    \includegraphics[width=\columnwidth]{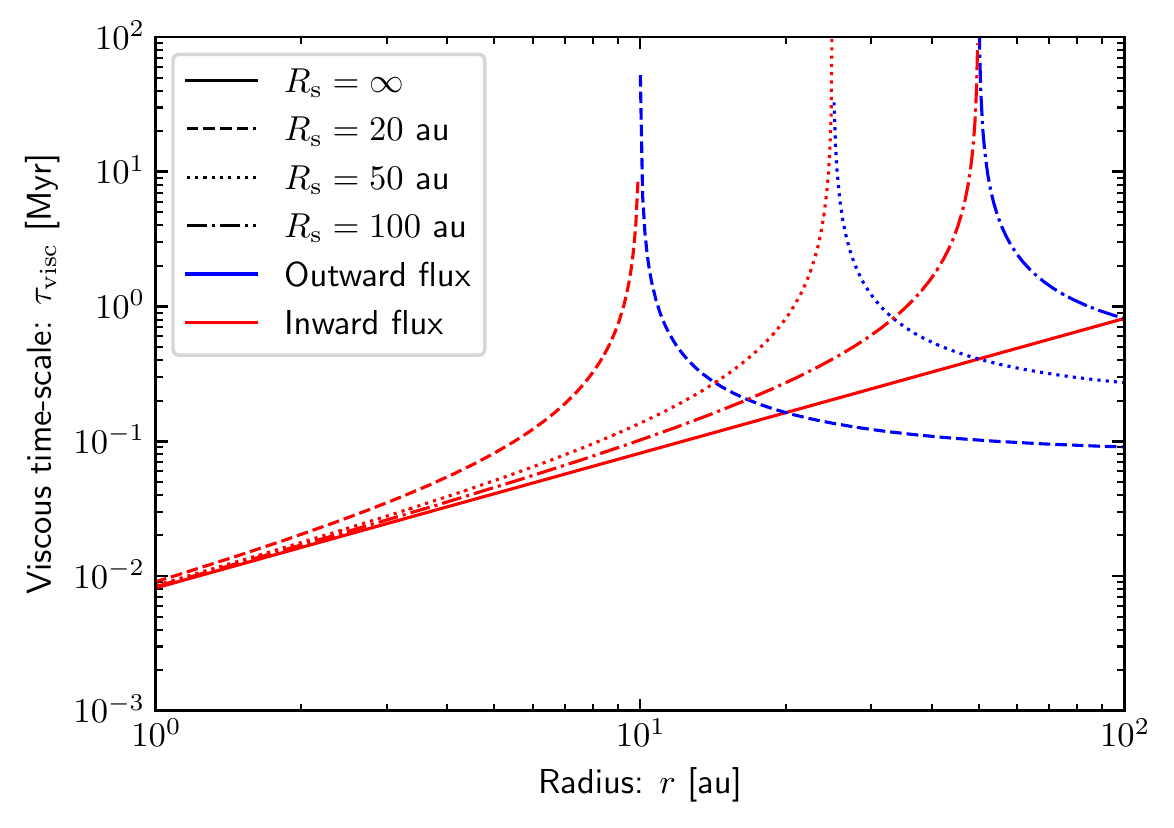}
    \caption{The viscous time-scale $\tau_\mathrm{visc} \equiv r/|v_r|$ as a function of radius for various disc surface density profiles. We show the outcome for various surface density profiles following equation~\ref{eq:Sigma0} but varying $R_\mathrm{s}$. We adopt a constant viscous $\alpha=3\cdot 10^{-3}$ and all other relevant parameters from our fiducial model (Table~\ref{tab:params}). Blue lines show outward flux ($v_r>0$) and red lines inward flux ($v_r<0$). }
    \label{fig:migtime}
\end{figure}

\subsubsection{Final masses}
In Figure~\ref{fig:outcome_exp} we explore how the initial semi-major axis $a_\mathrm{p,0}$ and external FUV flux $F_\mathrm{FUV}$ alter the final planet mass $M_\mathrm{p,f}$ (left panel) and semi-major axis $a_\mathrm{p,f}$ (right panel). We find that the final planet mass is dependent on $F_\mathrm{FUV}$ for all $a_\mathrm{p,0}$. This includes even close to the central star ($a_\mathrm{p,0}\lesssim 10$~au) , where externally driven winds are inefficient at driving mass-loss. This is because for a sufficiently large viscous $\alpha$, these winds are capable of shortening the inner disc lifetime by starving it of material that is redistributed from further out \citep[e.g.][]{Cla07}. Thus the masses of inner planets are limited in the time they are able to accrete gas. This is somewhat dependent on the formation time $\tau_\mathrm{form}$ and disc viscosity, as discussed in Sections~\ref{sec:form_time}~and~\ref{sec:visc} respectively. For our chosen parameters, the runaway growth phase is completely interrupted $F_\mathrm{FUV}\gtrsim 1000\,G_0$, yielding a population of low mass planets.



We find a dip in the masses of planets with $a_\mathrm{p,0} \approx 40$~au at $F_\mathrm{FUV}=10\,G_0$, decreasing to $a_\mathrm{p,0} \approx 10$~au for larger $F_\mathrm{FUV}\approx 2000\,G_0$, as seen in the left panel of Figure~\ref{fig:outcome_exp}. This is a consequence of the viscous mass flux in the disc. The local gas velocity due to viscous diffusion is:
\begin{equation}
    v_r = \frac{-3}{\Sigma r^{1/2}} \partial_r \left(\nu \Sigma r^{1/2}\right),
\end{equation} which also sets the type II migration in the disc-dominated limit \citet[][see also \citealt{Syer95}]{Armitage07}. Thus the rate of growth and the rate of migration is dependent on the shape of the surface density profile. We show the viscous time-scale in Figure~\ref{fig:migtime}, varying the scale radius $R_\mathrm{s}$. The transition between inwards and outwards mass flux is dependent on where the surface density profile steepens.

{To apply this intuition to our results, we can examine the trend in planet masses. We find a dip in planet mass at an initial semi-major axis that decreases with increasing $F_\mathrm{FUV}$. We similarly expect the outer disc radius that balances viscous outwards flux with the wind mass-loss rate to decrease with increasing $F_\mathrm{FUV}$ \citep{Winter20b, Hasegawa21}, and thus the transition between inwards and outwards flux also decreases. Hence at some intermediate radius, the viscous flux towards the planet is suppressed and the early growth (before a gap has opened up) is much slower than it would otherwise be. This dip exists because of the transition between inwards and outwards mass flux, and may therefore not be expected if angular momentum is transported magnetically \citep[e.g.][]{Tabone21}.}



\subsubsection{Migration}


The influence of external irradiation on migration is profound across all initial semi-major axes (right panel of Figure~\ref{fig:outcome_exp}). Even for $F_\mathrm{FUV} \sim 50\, G_0$, the degree to which the planet can migrate inwards is strongly suppressed due to the reduced disc surface density. This influences the type II migration stage; outside-in depletion means that the torque on the planet exerted by the outer disc becomes reduced with respect to the torque by the inner disc. 

For $F_\mathrm{FUV}\gtrsim 300\, G_0$, we find that inward migration is strongly suppressed across all $a_\mathrm{p,0}$. Even some marginal outwards migration is possible due to the removal of the outer disc material. At large separations and strong FUV fluxes the outer disc can be completely removed by the external wind. Once this material is removed, the disc wind becomes less efficient at removing material that is inside the gap opened by the planet due to the radius dependence \citep[e.g.][]{Haw18b}. The disc is therefore able to viscously expand, which also drives some outwards migration, although this expansion is effectively limited by the initial gap width. For intermediate fluxes, we therefore expect an accumulation of relatively low mass planets at the outer disc edge.

{The non-monotonic behaviour in the net migration of the planet in the right panel of Figure~\ref{fig:outcome_exp} can be understood in a similar way to the planet masses. At some specific $a_\mathrm{p}$, both the accretion onto the planet and the type II migration rate are strongly suppressed. This separation can be extremely localised (see Figure~\ref{fig:migtime}), such that if the planet is trapped in that location at the wrong time in its evolution it can undergo very little migration or growth.}

We have demonstrated that, for our fiducial model, external {photoevaporation} has a dramatic influence on the expected properties of planetary systems. In the following we investigate the influence of some of our chosen parameters on our results.

\subsection{Varying the core formation time}
\label{sec:form_time}
\begin{figure}
    \centering
    \includegraphics[width=\columnwidth]{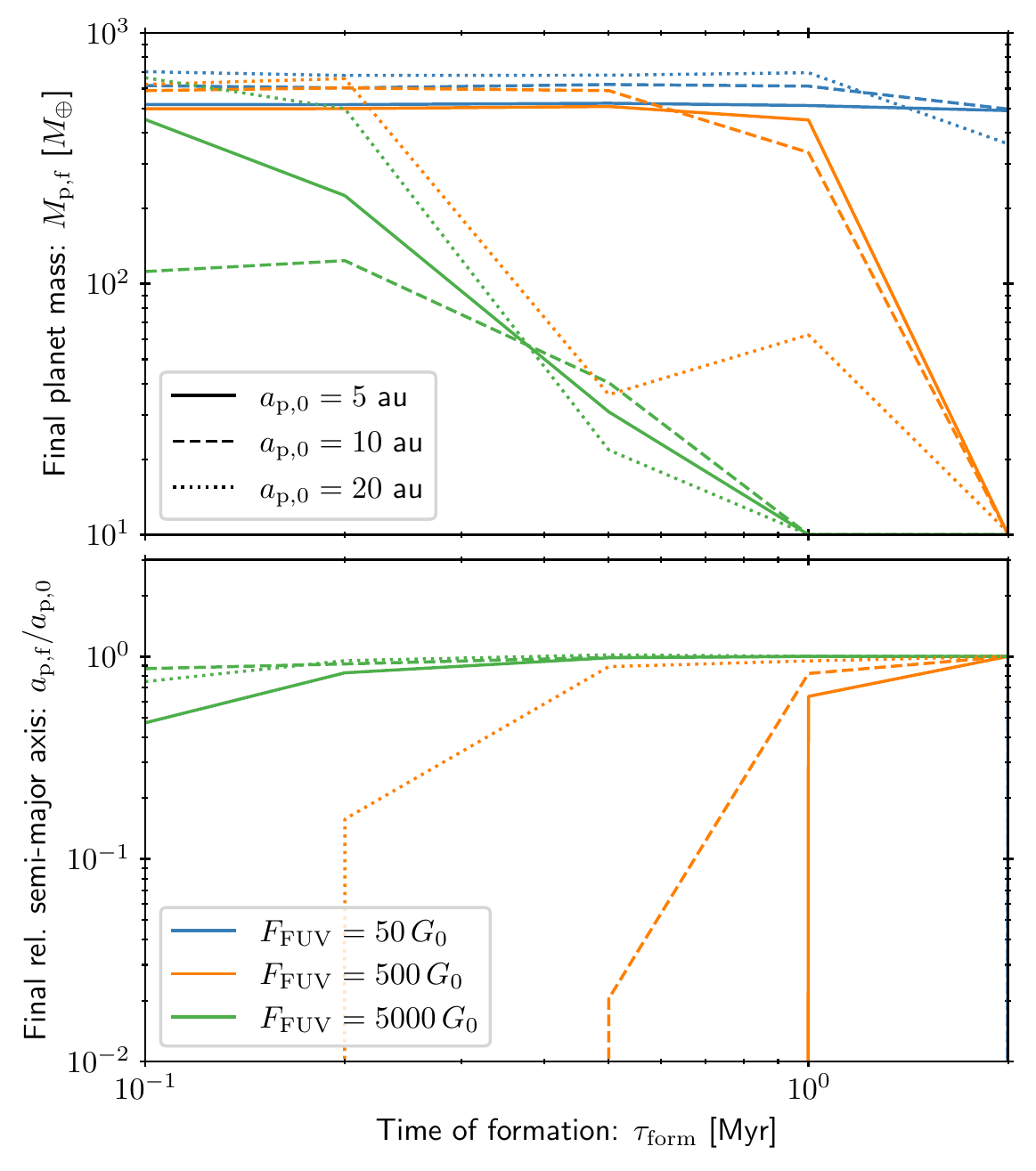}
    \caption{Variation of the final planet mass (top panel) and semi-major axis (bottom panel) in our fiducial model, but varying as a function of the formation time $\tau_\mathrm{form}$ and three different initial semi-major axes $a_\mathrm{p,0}=5$, $10$ and $20$~au (solid, dashed and dotted lines respectively). The colours of the lines indicate the external FUV flux $F_\mathrm{FUV} = 50$, $500$ and $5000 \, G_0$ (blue, {orange} and green respectively). The remaining parameters are as shown in Table~\ref{tab:params}. {Blue lines are not visible in the bottom panel because the planet migrates to $a_\mathrm{p}< 0.15$~au for all tested formation times. }  }
    \label{fig:tau_form}
\end{figure}

In Figure~\ref{fig:tau_form} we show the result of varying the formation time $\tau_\mathrm{form}$ of the $10 \, M_\oplus$ planet core in the inner disc ($a_\mathrm{p,0} = 5{-}20$~au). {A planet with smaller $\tau_\mathrm{form}$ has more time to grow and migrate, such that one expects lower masses and greater final separations for large $\tau_\mathrm{form}$. However, if $\tau_\mathrm{form}$ is too small then then planet can undergo rapid migration and accrete onto the central star. This results in optimum formation times for planet growth and survival. This optimum time decreases with increasing $a_\mathrm{p,0}$ and $F_\mathrm{FUV}$, which is because the externally-induced surface density suppression also occurs earlier. The formation time remains a significant uncertainty in our model, which may be addressed by modelling the initial formation stages.}



\subsection{Varying the disc viscosity}
\label{sec:visc}

\begin{figure}
    \centering
    \includegraphics[width=\columnwidth]{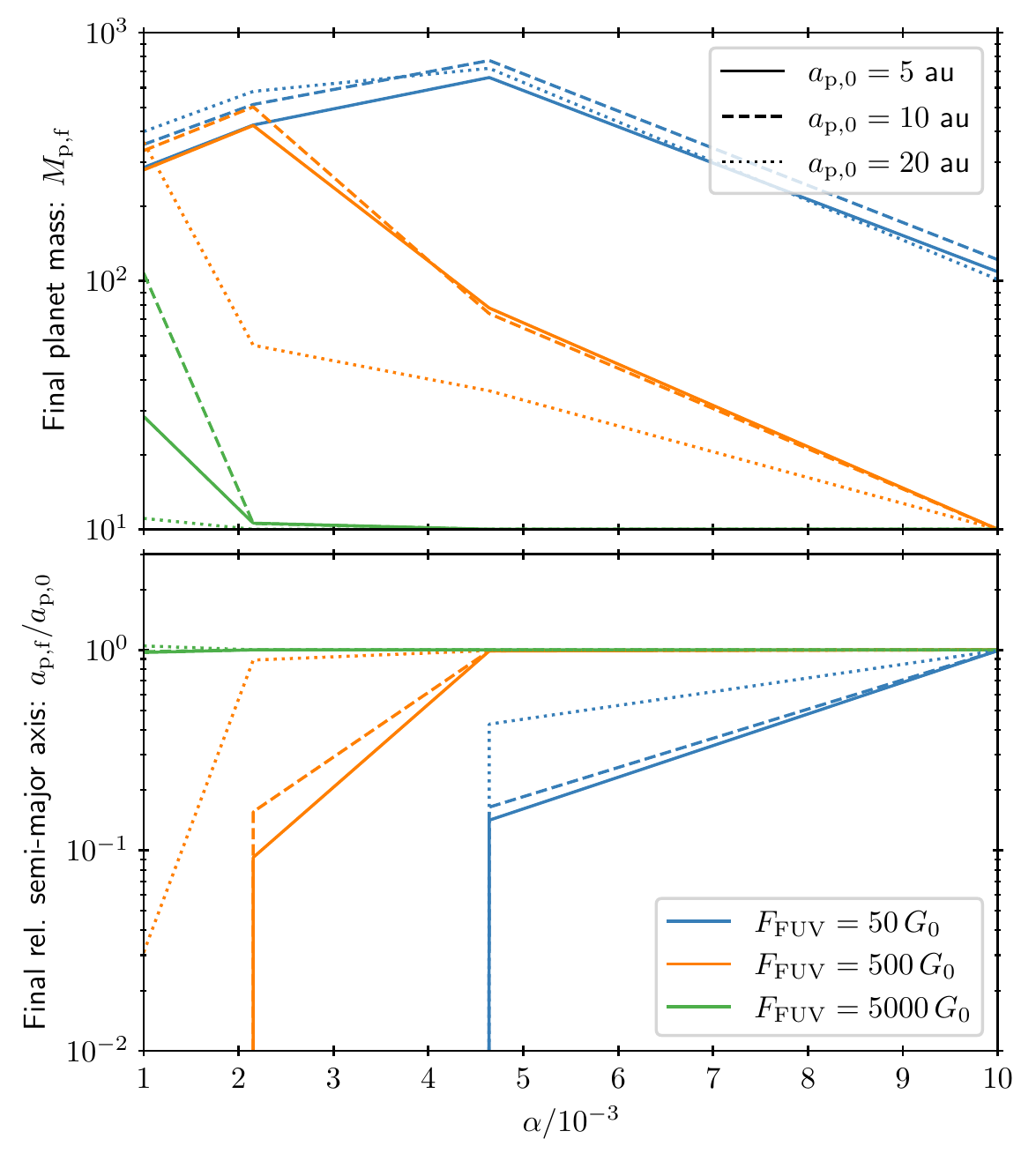}
    \caption{As in Figure~\ref{fig:tau_form}, but varying the viscous $\alpha$ rather than $\tau_\mathrm{form}$, which is fixed here to be  $\tau_\mathrm{form}=1$~Myr.}
    \label{fig:alpha}
\end{figure}

{In Figure~\ref{fig:alpha} we show the outcome of varying the viscous $\alpha$ in our model. Our results can be shortly summarised in that a low $\alpha$ results in accretion of the planet onto the central star. If the viscosity is high then the disc lifetime can be short, which thus reduces the time-scale for which the planet can accrete and migrate.}

{Large $\alpha$ is at odds with the apparent limits on turbulent viscosity from dust rings in ALMA observations, which suggest much lower $\alpha \lesssim 10^{-4}$ \citep{Pinte16}. However, we have chosen $\alpha=3\cdot 10^{-3}$ to retain the observed disc lifetimes and accretion rates, as discussed in Section~\ref{sec:int_wind}. If angular momentum is removed from the disc by magnetic winds then this should still lead to the growth of the planet as approximated by our $\alpha = 3\cdot  10^{-3}$ viscous model. This somewhat justifies our choice, but requires validation with models for magnetically mediated angular momentum transport \citep{Tabone21a}.}  



A caveat here is that we have not altered the mass-loss rate in the wind, although for higher viscosity a lower $\dot{M}_\mathrm{int}$ would be compatible with observed disc lifetimes (see Figure~\ref{fig:Mdot_acc}). This can contribute to halting inward migration \citep{Matsuyama03, Veras04}.

\subsection{Varying the stellar host mass}
\label{sec:mst}

\begin{figure}
    \centering
    \includegraphics[width=\columnwidth]{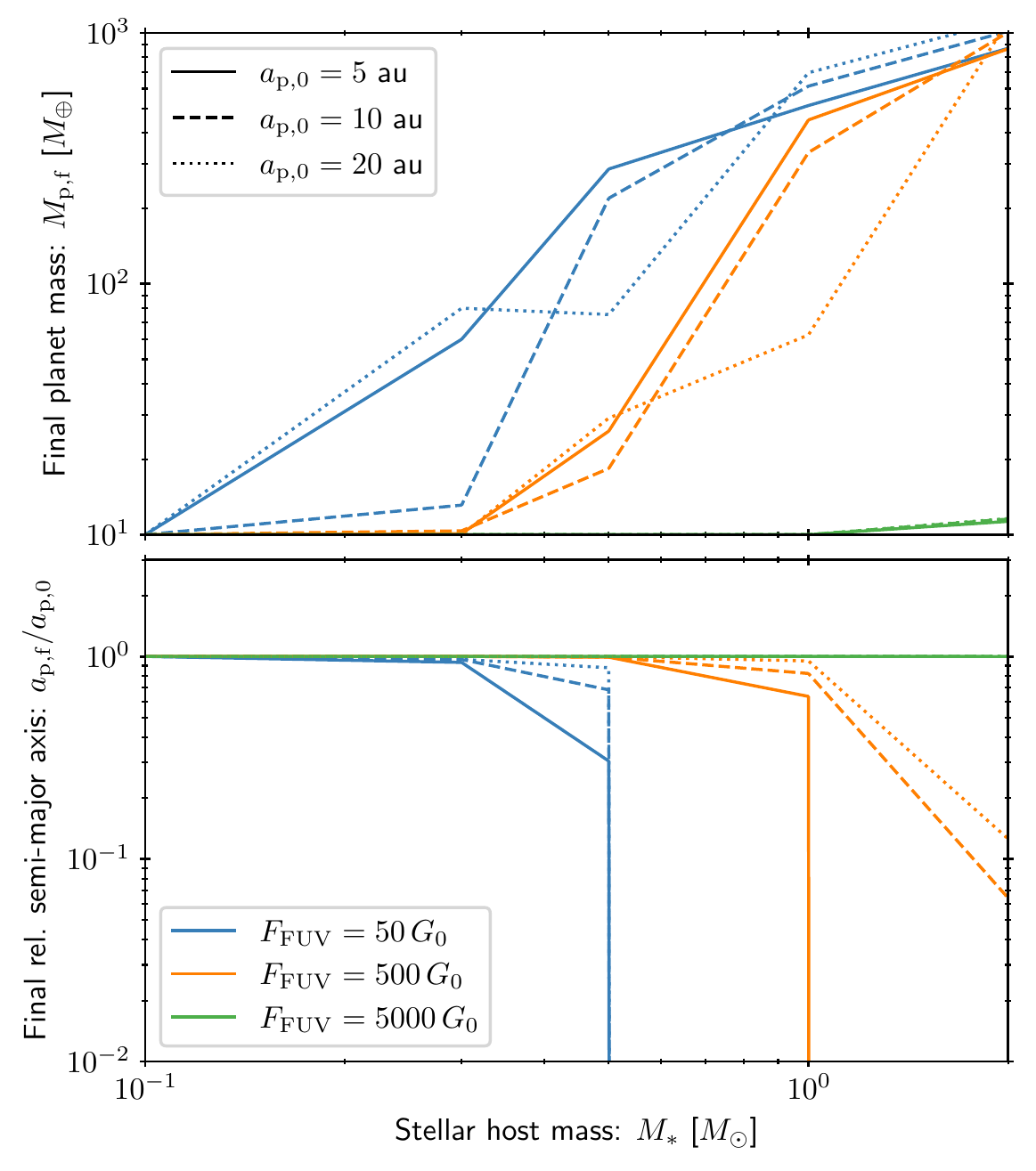}
    \caption{As in Figure~\ref{fig:tau_form}, but varying the stellar mass $M_*$ rather than $\tau_\mathrm{form}$, which is fixed here to be  $\tau_\mathrm{form}=1$~Myr.}
    \label{fig:mst}
\end{figure}
We now explore the dependence of the model outcomes on stellar mass. For this purpose, it is necessary to assume some scaling of the initial conditions with stellar mass $M_*$. We will simply assume that the initial surface density $\Sigma_0 \propto M_*$, and $L_* \propto M_*^{3.5}$, although the latter only marginally influences the assumed sound speed.

{The total mass-loss rate in the in the photoevaporative wind remains uncertain from an empirical perspective \citep[see][]{Ercolano17}.} Here we take an empirical and statistical approach to infer an approximate scaling with stellar mass. Observed accretion rates scale as $\dot{M}_\mathrm{acc} \propto M_*^\beta$ for $\beta\approx 2$ if a single power-law relationship is assumed \citep{Manara17}. {Since accretion rates of stars with discs undergoing viscous evolution with a fixed $\alpha$ drop geometrically, such stars should preferentially have accretion rates close to the transition to rapid inside-out depletion. Hence,} the mass-loss rate in the wind should be approximately proportional to the accretion rate at which a photoevaporative gap is opened up. We will therefore assume $\dot{M}_\mathrm{int} \propto M_*^2$. {This is steeper than implied by the numerical experiments of \citep{Picogna21}. However, adopting a shallower scaling does not influence our conclusions (see discussion in Section~\ref{sec:caveats}).}

The mass-loss rate in the external wind as a function of host star mass is determined directly by the outcome of the FRIED grid \citep{Haw18b}. 

The outcome of this experiment is shown in Figure~\ref{fig:mst}. We find that the final planet mass (top panel) is a strong, almost monotonically increasing function of the stellar host mass ($M_*< 2 \,M_\odot$). This is broadly consistent with the finding that massive planets have higher occurrence rates around higher mass stars up to $M_* = 1.9\, M_\odot$ \citep{Johnson10, Reffert15}. In order to form planets of masses $M_\mathrm{p}\gtrsim 50\, M_\oplus$ around a star of mass $M_*\lesssim 0.5\,M_\odot$, the external radiation field must be $F_\mathrm{FUV} \lesssim 500\, G_0$ -- i.e. lower than the solar neighbourhood average \citep{Fat08}. Even in this case, inward migration becomes inefficient. Because planets at greater orbital distance are harder to detect, this may further contribute to the apparent absence of massive planets around low mass stars.

\subsection{Comparison to current prescription }

\begin{figure}
    \centering
    \includegraphics[width=\columnwidth]{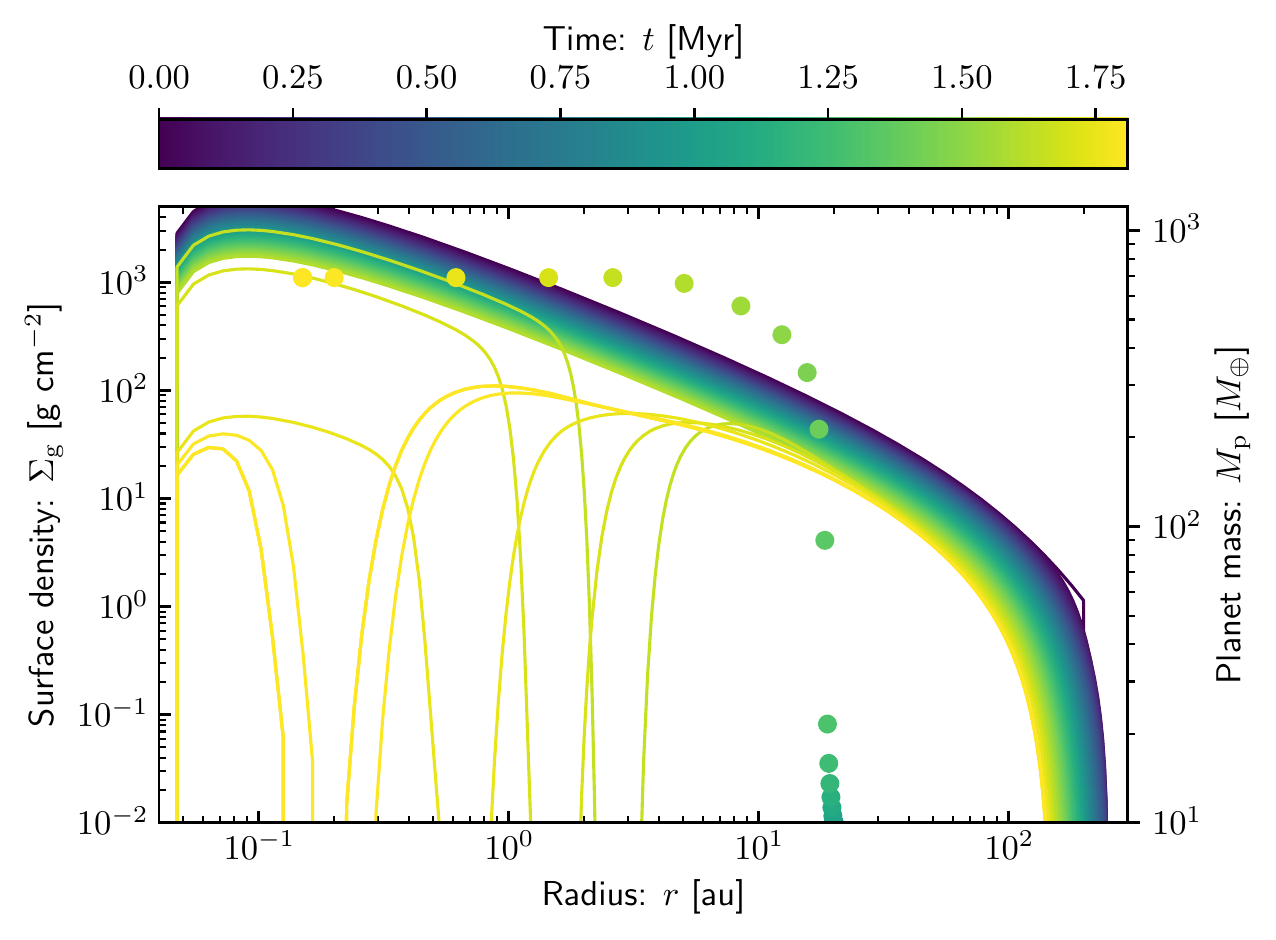}
    \caption{Similar to Figure~\ref{fig:sd_evol}, except with the prescription for external {photoevaporation} implemented by \citet{Emsenhuber21}. See text for details. }
    \label{fig:E20_comp}
\end{figure}

{We consider how the outcome of our model compares to the prescription for external disc depletion adopted by state-of-the-art population synthesis models. \citet{Emsenhuber21} adopts a prescription for the externally driven mass-loss that is a constant change in surface density: }
\begin{equation}
   \dot{ \Sigma}_\mathrm{ext, E21} = \begin{cases} 0 &\qquad r<R_\mathrm{crit}\\
  \frac{ \dot{M}_\mathrm{ext, max}}{2\pi (R_\mathrm{max}^2- R_\mathrm{crit}^2)} &\qquad r\geq R_\mathrm{crit}\\
   \end{cases}.
\end{equation}{Here  $\dot{M}_\mathrm{ext, max}=6.42\times 10^{-7} \, M_\odot$~yr$^{-1}$ is the total mass-loss rate if the disc extended out to $R_\mathrm{max}$, with $R_\mathrm{max}=1000$~au as for \citet{Emsenhuber21}, and the critical radius $R_\mathrm{crit}=R_\mathrm{launch}$. This model therefore {imposes} constant depletion of the surface density as a function of radius, rather than an outside-in depletion that should be expected theoretically \citep[e.g.][]{Johnstone98}.}

We show the outcome of this simplified version of the external wind with our model in Figure~\ref{fig:E20_comp}. In this case, the disc is not strongly depleted by the external wind before the planet has migrated inwards and accreted into its host star. This outcome is comparable to our very low $F_\mathrm{FUV}\lesssim 100 \, G_0$ models. Such a flux is far lower than the solar neighbourhood average \citep{Fat08, Winter20a}. In addition, the outer disc radius in the simplified model is no longer set by the balance of external mass-loss and viscous outwards angular momentum redistribution. {The slight contraction of the disc seen in Figure~\ref{fig:E20_comp} is due to the initially lower surface density in the outer disc.} This has consequences for the mass-flux throughout the disc, and would not result in the dip in planet masses we find in Figure~\ref{fig:outcome_exp}. We conclude that correctly modelling external photoevaporative mass-loss is an important consideration for planet population synthesis studies. 

\subsection{Implications for observations and theory}
\label{sec:comp_obs}

\begin{figure*}
    \centering
    \includegraphics[width=\textwidth]{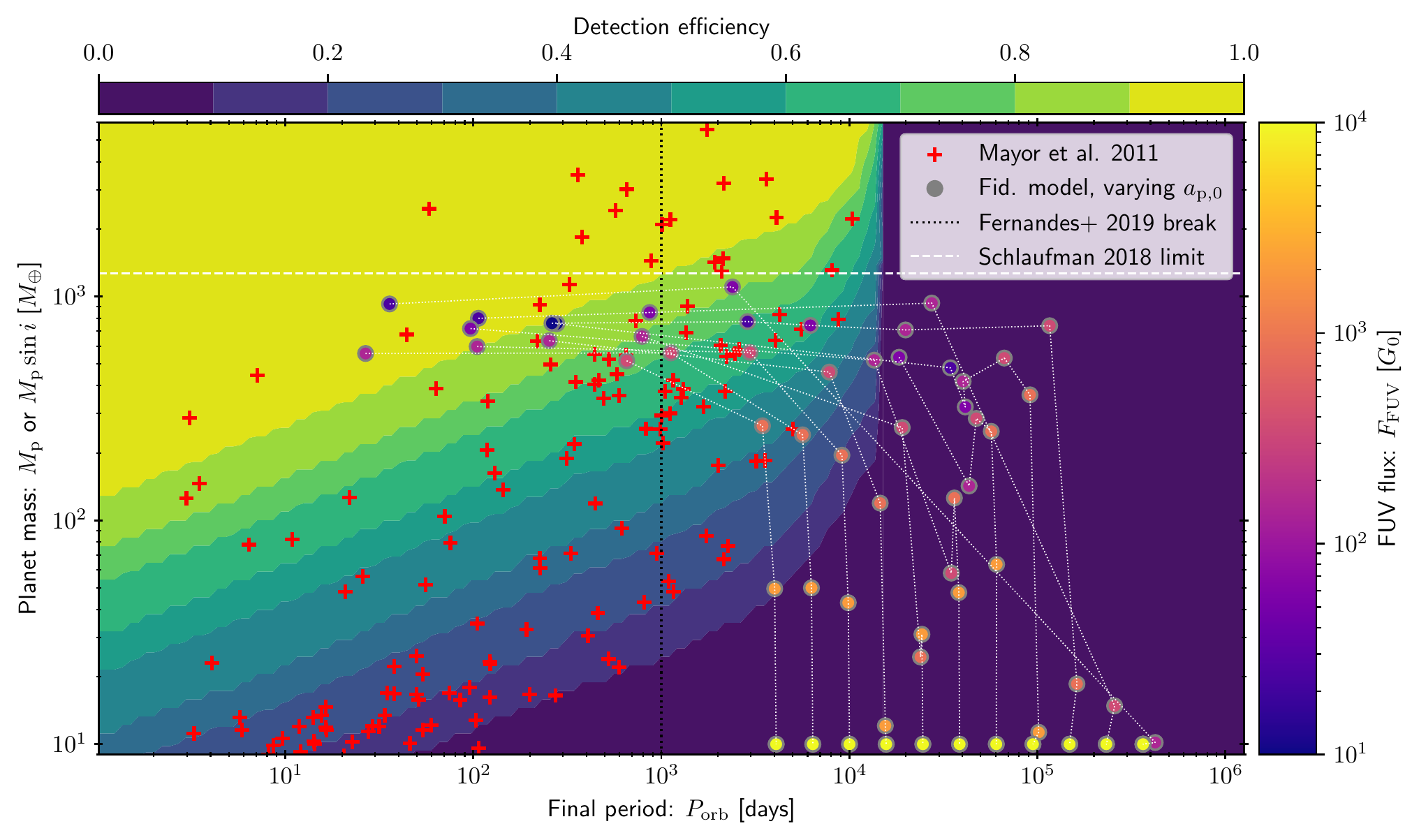}
    \caption{The radial velocity planet discoveries from the HARPS survey \citep[red crosses][]{Mayor11}, compared with the outcome of our fiducial model (circles) only varying the initial semi-major axis $a_\mathrm{p,0}$ and FUV flux $F_\mathrm{FUV}$ (right colour bar) across the grid points shown in Figure~\ref{fig:outcome_exp}. Points in our models are connected by faint white lines connecting points of fixed  $a_\mathrm{p,0}$. The occurrence rate peak in orbital period $P_\mathrm{orb}$ space as inferred by \citet{Fernandes19} is shown as a vertical dotted black line, with the detection efficiency shown by colours indicated in the top colour bar. The horizontal dashed white line shows the approximate threshold above which sub-stellar companions may form by an alternative mechanism \citep{Schlaufman18}. Detected exoplanet masses are measured in projection, $M_\mathrm{p}\sin i$ for inclination $i$. }
    \label{fig:comp_ob}
\end{figure*}

We now consider what our results mean in the context of observed and theoretical planet properties. Given the simplicity of our models, we do not attempt a population synthesis or similar. Instead, we focus on the robust outcomes of the models we have presented. We suggest how they may be relevant for the modern exoplanet and planet formation communities, and where they may be developed in future.

\subsubsection{Theoretical models}

 Most obviously, in terms of theoretical expected exoplanet properties, we have demonstrated that external FUV irradiation has a strong influence on the final masses and semi-major axes of planets. In particular, for population efforts in which type II migration plays a significant role \citep[e.g.][]{Alibert05, Alibert13, Emsenhuber21} external photoevaporation is an important addition to the physics of the models that cannot be neglected. Ignoring interstellar extinction, typical FUV fluxes experienced by discs are $F_\mathrm{FUV} \sim 3000 \, G_0$ \citep{Fat08, Winter20a}, which is easily sufficient to strongly influence the planet population. 
 
 In terms of realistic FUV flux exposure for young star-disc systems, \citet{Ali19} performed simulations of feedback form a single massive star on the forming stars and gas. They found that objects tend to only be shielded from the FUV by residual gas in the star forming environment for $<0.5$~Myr. This time-scale may reduce further in regions with multiple massive stars, where gas ejection via feedback is faster and it is geometrically more challenging to shield a star from the FUV flux. Thus the interstellar extinction, while an important consideration early on, is unlikely to mitigate the role of FUV fields in sculpting the massive planet population. 
 
 Perhaps the most important absence in our models is a treatment of the growth and migration of solids in the disc. Previous studies have indicated that external photoevaporation may enhance dust to gas ratios to induce streaming instability in irradiated discs \citep{Throop05}. In addition, grain growth, migration and depletion can all be influenced by external irradiation \citep{Sellek20}. Coupling this understanding with planet formation models remains an important step in interpreting exoplanet populations via modelling. 

\subsubsection{Comparison to radial velocity planet discoveries}

With regard to observational constraints, the prevalence of FUV sculpted exoplanet populations offers an alternative explanation for some empirical trends obtained from exoplanet surveys. For example, in Section~\ref{sec:mst} we discuss the suppression of giant planet formation around low mass stars consistent with the inferred suppressed occurrence rates \citep[e.g.][]{Johnson10}. 

We also consider the final planet properties obtained from our fiducial model in comparison with the discoveries of the HARPS radial velocity survey \citep{Mayor11} in Figure~\ref{fig:comp_ob}. \citet{Fernandes19} used the detection efficiency of that survey (indicated by the top colour bar) to infer the existence of a peak in the massive planet occurrence rates at $P_\mathrm{orb} \sim 10^3$~days. This was interpreted as evidence of an accumulation of planets at the snow line. However, whether this is physically expected is sensitively dependent on assumptions about planet growth and migration \citep[e.g.][]{Mordasini18, Ida18}. 

An alternative explanation becomes clear from the distribution of planets in $a_\mathrm{p}{-}M_\mathrm{p}$ space in our model. We show that increasing the FUV flux can produce an anti-correlation between $M_\mathrm{p}$ and $a_\mathrm{p}$. If this anti-correlation is present in the exoplanet population, then due to the declining detection efficiency with decreasing planet mass one would expect to detect fewer planets than expected with increasing $P_\mathrm{orb}$. Note that there is no reason to expect this finding to apply to planets with masses $M_\mathrm{p}\gtrsim 4\,M_\mathrm{J}$, which may not have formed by core accretion \citep{Schlaufman18}. Whether or not this explanation adequately explains the observed peak in occurrence rate requires a combination of longer baseline, sensitive radial velocity surveys and a development of the models presented here that treats the earlier stages of planet formation. Of course, other origins for the apparent pile-up, such as internal photoevaporative winds, remain possible \citep{Alexander12b, Ercolano15, Coleman16}.

{From Figure~\ref{fig:comp_ob}, we can also estimate the FUV field strengths required to produce the short period massive planets. Those close to the peak in occurrence rate for planets with mass $M_\mathrm{p} >30\,M_\oplus$ inferred by \citet{Fernandes19} would be formed in environments less than a few $100\, G_0$. This is lower than in the typical birth environment for stars in solar neighbourhood \citep{Fat08, Winter20a}. However, the occurrence rate for such planets orbiting at periods $10^2{-}10^4$~days is $\sim 10{-}20$~percent, indicating that the majority of stars may host lower mass planets that are not detected by current RV surveys. Future efforts to combine our prescription with population synthesis may reveal whether these occurrence rates are consistent with the expected distribution of FUV fluxes in stellar birth environments.  }

Finally, given that we have demonstrated that gas accretion can be interrupted or completely suppressed, we may expect that envelope accretion is less common in dense environments. Empirically, \citet{Kruijssen20} found that stars that occupy overdensities in position-velocity space host planets that are less likely to retain a gaseous envelope. Whether this is due to the formation environment remains an open question. Our findings would suggest that external photoevaporation is a feasible mechanism to suppress the formation of planets with massive gaseous envelopes. 

\subsubsection{Protoplanetary disc demographics}

{The relevance of our models depends on the degree to which observed protoplanetary discs are sculpted by environment. From a theoretical perspective, \citet{Coleman22} demonstrate how the balance of external and internal photoevaporation affect the evolution of discs. The authors suggest that external photoevaporation can dominate mass-loss even at low FUV fluxes, while inside-out dispersal driven by internal photoevaporation is less prevalent in high FUV environments. In general, empirical studies aimed at demonstrating how environment plays a role in disc evolution are hindered by challenges in accurately determining stellar ages \citep[e.g.][]{Bell13} and possibly complex star formation and dynamical histories \citep[e.g.][]{Hil98,Get14,Bec17,Winter19b}. Nonetheless, it is clear that in at least some very massive star forming regions with several O stars, external radiation shortens the inner disc lifetime \citep[][]{Stolte04, Fang12,Gua16}. Both the total dust content \citep{Eis18, vTer19} and the outer gas radii \citep{Boyden20} of discs in the Orion Nebula cluster, which is the nearest intermediate mass star forming region, also appear to be depleted by this process.  }

{A full comparison of the observed protoplanetary disc demographics to our model outcomes is beyond the scope of this work. Such a comparison would require both a prescription for the dust evolution and a population synthesis approach including stellar dynamical histories \citep[see, for example,][]{Winter19b, Concha-Ramirez19, Sellek20, Qiao22}. However, evidence of rapid mass-loss from the proplyds in the moderate FUV environments of NGC 1977 \citep{Kim16} and NGC 2024 \citep{Haworth21}, as well as depleted dust masses close to $\sigma$ Ori \citep{Ans17} imply that discs are somewhat influenced in such environments.    }

\subsection{Caveats and future developments}
\label{sec:caveats}
We have here considered only a very simple model for the growth and migration of massive planets within the protoplanetary disc. We adopt a simplified approach in order to be able to clearly identify the role of external photoevaporation versus a plethora of other processes that may be relevant for the formation of planets. A non-exhaustive list of some major simplifications are as follows:
\begin{itemize}
    \item \textit{Multiple planets:} We have considered only a single planet evolving in the disc in order to isolate the role that external depletion plays on massive planet formation. Interactions between multiple forming planets may have an influence on the final planet orbit and mass \citep[e.g.][]{Alibert13}.   
    \item \textit{Early growth:} Our initial conditions have assumed that the core growth has already occurred, and in doing so we ignore all of the early stage physics of planetesimal and planet embryo formation \citep[e.g.][]{Voelkel21, Coleman21}. These uncertainties are partially parameterised by the time $\tau_\mathrm{form}$ at which our massive planet core is injected. However, $\tau_\mathrm{form}$ is probably dependent on the location in the disc. Further, early growth stages may influence the local disc properties at time $\tau_\mathrm{form}$, which could in turn alter the outcome of models. 
    \item \textit{Type I migration:} {Here we have followed \citet{Paardekooper11} for the type I migration rate, but include strong suppression of the migration rate by a factor $10^{-2}$. Population synthesis studies frequently adopt some slow down of the type I migration rate \citep[see discussion by][]{Benz14}. It is not understood what the true type I migration rate should be; MHD turbulence \citep{Nelson04} or thermal torques \citep{Benitez15} may suppress inwards migration. However, it is unclear where and for how long such influences might suppress migration \citep{Guilera19, Guilera21}.}
    \item \textit{Disc viscosity:} We adopt a viscosity parameter of $\alpha=3\cdot 10^{-3}$, which yields reasonable accretion rates and disc lifetimes in our model. However, if angular momentum transport in the disc is mediated by magnetohydrodynamic effects then the disc evolution may be considerably different \citep[e.g.][]{Tabone21}. {Adopting such a model instead would influence our findings to some degree. However, a comparable inward flux of material from the outer disc should sustain similar growth rates for inner planets.}
    \item {\textit{Stellar mass dependences:} We have adopted an internal wind mass-loss rate scaling $\dot{M}_\mathrm{int} \propto M_*^2$, which is steeper than found in numerical experiments \citep{Owen12,Picogna21}. This choice was made to approximately reproduce the observed scaling of accretion rates in low mass star forming regions \citep{Manara17}, although a shallower relationship may be physically appropriate. Our findings in Section~\ref{sec:mst} show that the final planet mass (semi-major axis) increases (decreases) monotonically and steeply with stellar host mass. This is the result of a longer period of growth and migration for higher mass host stars with respect to lower mass host stars. Reducing the dependence of $\dot{M}_\mathrm{int}$ on $M_*$ acts to further steepen this relationship, leaving our conclusions unchanged. Nonetheless, the true dependence of $\dot{M}_\mathrm{int}$ on $M_*$ may combine with variations in other properties (such as formation time of the planet core) to alter how giant planet formation in irradiated environments depends on external irradiation.}
    \item {\textit{Flow across the gap:} Studies have previously shown that the flow of material across a planet-carved gap and efficiency of planetary accretion is important for setting the migration rate of a giant planet \citep{Ale12, Monsch21b}. We have adopted a fitting prescription here following previous authors (see Section~\ref{sec:accretion}), however this is not necessarily an adequate substitute for full two or three dimensional simulations in this context.}
\end{itemize}

Thus our results should be regarded a first estimate on the potential for external photoevaporation to influence planet populations throughout the disc. Numerous avenues to deal with the above concerns should be explored in future. 

\section{Conclusions}
\label{sec:conclusions}

In this work, we have investigated the role of external FUV fields in regulating the growth and migration of massive planets during their formation. We have considered a simple one dimensional disc-planet evolution model, including prescriptions for viscous evolution, internal and external photoevaporation, and type I and II migration. Our major findings are summarised as follows:

\begin{enumerate}
    \item For FUV flux $F_\mathrm{FUV}\gtrsim 100\, G_0$, external photoevaporation has a significant influence on planet growth and migration regardless of the initial formation location and assumed formation parameters. In general, the influence of the external depletion is to reduce the final planet mass and suppress inward migration. Given that typically discs in the solar neighbourhood experience $F_\mathrm{FUV}\sim 3000\, G_0$ in their formation environment \citep{Fat08, Winter20a}, this is an important factor in planet population synthesis efforts.
    \item For inner planets, FUV irradiation results in an anti-correlation between the semi-major axis $a_\mathrm{p}$ and the planet mass $M_\mathrm{p}$. Given that the detection efficiency of radial velocity surveys decreases with increasing $a_\mathrm{p}$ and decreasing $M_\mathrm{p}$, this may contribute to the apparent peak in the occurrence of massive planets at orbital periods $P_\mathrm{orb}\sim 10^3$~days \citep{Fernandes19}.
    \item External photoevaporation is more efficient at suppressing giant planet growth and migration around lower mass stars. For stars $M_*\lesssim 0.5\,M_\odot$, giant planet formation is difficult to achieve for $F_\mathrm{FUV} \gtrsim 500\, G_0$, which is typical for discs evolving in the solar neighbourhood. Such planets also undergo little or no inward migration. This may help to explain the low occurrence rates for giant planets around low mass stars \citep{Johnson10}. 
    \item Our models show that the correct prescription for external photoevaporation and dependence on the external FUV flux is important for planet population synthesis efforts. The state-of-the-art population synthesis efforts underestimate the impact of external disc depletion on the final planet population.
\end{enumerate}

These findings are the outcome of simple models, and apply only to giant planet formation once a massive core has already formed. They are intended to complement previous efforts highlighting the importance of external photoevaporation for the early evolution of solids in protoplanetary discs \citep[e.g.][]{Throop05, Miotello12, Sellek20}. These calculations motivate more detailed future simulations.

\section*{Data availability}

The code and model outcomes used to produce figures in this work are available upon reasonable request to the corresponding author.

\section*{Software}

{We thank the contributors for making \textsc{Matplotlib} \citep{Hunter07},
\textsc{Numpy} \citep{Harris20} and \textsc{Scipy} \citep{Virtanen20}, of which we made use in this work, public. }

\section*{Acknowledgements}

{We sincerely thank the anonymous referee for their careful reading that significantly improved the robustness and clarity of this manuscript.} AJW thanks Richard Alexander for useful discussion regarding the numerical treatment of type II migration. AJW acknowledges funding from an Alexander von Humboldt Stiftung Postdoctoral Research Fellowship. TJH is funded by a Royal Society Dorothy Hodgkin Fellowship. GALC was funded by the Leverhulme Trust through grant RPG-2018-418. SN acknowledges the funding from the
UK Science and Technologies Facilities Council, grant No. ST/S000453/1.




\bibliographystyle{mnras}
\bibliography{bdbib} 

\begin{thebibliography}{}
\makeatletter
\relax
\def\mn@urlcharsother{\let\do\@makeother \do\$\do\&\do\#\do\^\do\_\do\%\do\~}
\def\mn@doi{\begingroup\mn@urlcharsother \@ifnextchar [ {\mn@doi@}
  {\mn@doi@[]}}
\def\mn@doi@[#1]#2{\def\@tempa{#1}\ifx\@tempa\@empty \href
  {http://dx.doi.org/#2} {doi:#2}\else \href {http://dx.doi.org/#2} {#1}\fi
  \endgroup}
\def\mn@eprint#1#2{\mn@eprint@#1:#2::\@nil}
\def\mn@eprint@arXiv#1{\href {http://arxiv.org/abs/#1} {{\tt arXiv:#1}}}
\def\mn@eprint@dblp#1{\href {http://dblp.uni-trier.de/rec/bibtex/#1.xml}
  {dblp:#1}}
\def\mn@eprint@#1:#2:#3:#4\@nil{\def\@tempa {#1}\def\@tempb {#2}\def\@tempc
  {#3}\ifx \@tempc \@empty \let \@tempc \@tempb \let \@tempb \@tempa \fi \ifx
  \@tempb \@empty \def\@tempb {arXiv}\fi \@ifundefined
  {mn@eprint@\@tempb}{\@tempb:\@tempc}{\expandafter \expandafter \csname
  mn@eprint@\@tempb\endcsname \expandafter{\@tempc}}}

\bibitem[\protect\citeauthoryear{{ALMA Partnership} et~al.,}{{ALMA Partnership}
  et~al.}{2015}]{HLTau_15}
{ALMA Partnership} et~al., 2015, \mn@doi [\apjl] {10.1088/2041-8205/808/1/L3},
  \href {https://ui.adsabs.harvard.edu/abs/2015ApJ...808L...3A} {808, L3}

\bibitem[\protect\citeauthoryear{{Adamo} et~al.,}{{Adamo}
  et~al.}{2017}]{Adamo17}
{Adamo} A.,  et~al., 2017, \mn@doi [\apj] {10.3847/1538-4357/aa7132}, \href
  {https://ui.adsabs.harvard.edu/abs/2017ApJ...841..131A} {841, 131}

\bibitem[\protect\citeauthoryear{{Adams}, {Hollenbach}, {Laughlin}  \&
  {Gorti}}{{Adams} et~al.}{2004}]{Ada04}
{Adams} F.~C.,  {Hollenbach} D.,  {Laughlin} G.,   {Gorti} U.,  2004, \mn@doi
  [\apj] {10.1086/421989}, 611, 360

\bibitem[\protect\citeauthoryear{{Alexander}}{{Alexander}}{2012}]{Ale12}
{Alexander} R.,  2012, \mn@doi [\apjl] {10.1088/2041-8205/757/2/L29}, 757, L29

\bibitem[\protect\citeauthoryear{{Alexander} \& {Armitage}}{{Alexander} \&
  {Armitage}}{2009}]{Alexander09}
{Alexander} R.~D.,  {Armitage} P.~J.,  2009, \mn@doi [\apj]
  {10.1088/0004-637X/704/2/989}, \href
  {https://ui.adsabs.harvard.edu/abs/2009ApJ...704..989A} {704, 989}

\bibitem[\protect\citeauthoryear{{Alexander} \& {Pascucci}}{{Alexander} \&
  {Pascucci}}{2012}]{Alexander12b}
{Alexander} R.~D.,  {Pascucci} I.,  2012, \mn@doi [\mnras]
  {10.1111/j.1745-3933.2012.01243.x}, \href
  {https://ui.adsabs.harvard.edu/abs/2012MNRAS.422L..82A} {422, L82}

\bibitem[\protect\citeauthoryear{{Ali} \& {Harries}}{{Ali} \&
  {Harries}}{2019}]{Ali19}
{Ali} A.~A.,  {Harries} T.~J.,  2019, \mn@doi [\mnras] {10.1093/mnras/stz1673},
  \href {https://ui.adsabs.harvard.edu/abs/2019MNRAS.487.4890A} {487, 4890}

\bibitem[\protect\citeauthoryear{{Alibert}, {Mordasini}, {Benz}  \&
  {Winisdoerffer}}{{Alibert} et~al.}{2005}]{Alibert05}
{Alibert} Y.,  {Mordasini} C.,  {Benz} W.,   {Winisdoerffer} C.,  2005, \mn@doi
  [\aap] {10.1051/0004-6361:20042032}, \href
  {https://ui.adsabs.harvard.edu/abs/2005A&A...434..343A} {434, 343}

\bibitem[\protect\citeauthoryear{{Alibert}, {Carron}, {Fortier}, {Pfyffer},
  {Benz}, {Mordasini}  \& {Swoboda}}{{Alibert} et~al.}{2013}]{Alibert13}
{Alibert} Y.,  {Carron} F.,  {Fortier} A.,  {Pfyffer} S.,  {Benz} W.,
  {Mordasini} C.,   {Swoboda} D.,  2013, \mn@doi [\aap]
  {10.1051/0004-6361/201321690}, \href
  {https://ui.adsabs.harvard.edu/abs/2013A&A...558A.109A} {558, A109}

\bibitem[\protect\citeauthoryear{{Anderson}, {Adams}  \& {Calvet}}{{Anderson}
  et~al.}{2013}]{Ande13}
{Anderson} K.~R.,  {Adams} F.~C.,   {Calvet} N.,  2013, \mn@doi [\apj]
  {10.1088/0004-637X/774/1/9}, 774, 9

\bibitem[\protect\citeauthoryear{{Ansdell}, {Williams}, {Manara}, {Miotello},
  {Facchini}, {van der Marel}, {Testi}  \& {van Dishoeck}}{{Ansdell}
  et~al.}{2017}]{Ans17}
{Ansdell} M.,  {Williams} J.~P.,  {Manara} C.~F.,  {Miotello} A.,  {Facchini}
  S.,  {van der Marel} N.,  {Testi} L.,   {van Dishoeck} E.~F.,  2017, \mn@doi
  [\aj] {10.3847/1538-3881/aa69c0}, 153, 240

\bibitem[\protect\citeauthoryear{{Armitage}}{{Armitage}}{2007}]{Armitage07}
{Armitage} P.~J.,  2007, \mn@doi [\apj] {10.1086/519921}, \href
  {https://ui.adsabs.harvard.edu/abs/2007ApJ...665.1381A} {665, 1381}

\bibitem[\protect\citeauthoryear{{Bai} \& {Stone}}{{Bai} \&
  {Stone}}{2013}]{Bai13}
{Bai} X.-N.,  {Stone} J.~M.,  2013, \mn@doi [\apj]
  {10.1088/0004-637X/769/1/76}, \href
  {https://ui.adsabs.harvard.edu/abs/2013ApJ...769...76B} {769, 76}

\bibitem[\protect\citeauthoryear{{Ballabio}, {Alexander}  \&
  {Clarke}}{{Ballabio} et~al.}{2020}]{Ballabio20}
{Ballabio} G.,  {Alexander} R.~D.,   {Clarke} C.~J.,  2020, \mn@doi [\mnras]
  {10.1093/mnras/staa1767}, \href
  {https://ui.adsabs.harvard.edu/abs/2020MNRAS.496.2932B} {496, 2932}

\bibitem[\protect\citeauthoryear{{Barge} \& {Sommeria}}{{Barge} \&
  {Sommeria}}{1995}]{Barge95}
{Barge} P.,  {Sommeria} J.,  1995, \aap, \href
  {https://ui.adsabs.harvard.edu/abs/1995A&A...295L...1B} {295, L1}

\bibitem[\protect\citeauthoryear{{Baruteau} et~al.,}{{Baruteau}
  et~al.}{2014}]{Baruteau14}
{Baruteau} C.,  et~al., 2014, in {Beuther} H.,  {Klessen} R.~S.,  {Dullemond}
  C.~P.,   {Henning} T.,  eds, Protostars and Planets VI. p.~667 (\mn@eprint
  {arXiv} {1312.4293}), \mn@doi{10.2458/azu\_uapress\_9780816531240-ch029}

\bibitem[\protect\citeauthoryear{{Beccari} et~al.,}{{Beccari}
  et~al.}{2017}]{Bec17}
{Beccari} G.,  et~al., 2017, \mn@doi [\aap] {10.1051/0004-6361/201730432}, 604,
  A22

\bibitem[\protect\citeauthoryear{{Bell}, {Naylor}, {Mayne}, {Jeffries}  \&
  {Littlefair}}{{Bell} et~al.}{2013}]{Bell13}
{Bell} C. P.~M.,  {Naylor} T.,  {Mayne} N.~J.,  {Jeffries} R.~D.,
  {Littlefair} S.~P.,  2013, \mn@doi [\mnras] {10.1093/mnras/stt1075}, \href
  {https://ui.adsabs.harvard.edu/abs/2013MNRAS.434..806B} {434, 806}

\bibitem[\protect\citeauthoryear{{Ben{\'\i}tez-Llambay}, {Masset},
  {Koenigsberger}  \& {Szul{\'a}gyi}}{{Ben{\'\i}tez-Llambay}
  et~al.}{2015}]{Benitez15}
{Ben{\'\i}tez-Llambay} P.,  {Masset} F.,  {Koenigsberger} G.,   {Szul{\'a}gyi}
  J.,  2015, \mn@doi [\nat] {10.1038/nature14277}, \href
  {https://ui.adsabs.harvard.edu/abs/2015Natur.520...63B} {520, 63}

\bibitem[\protect\citeauthoryear{{Benz}, {Ida}, {Alibert}, {Lin}  \&
  {Mordasini}}{{Benz} et~al.}{2014}]{Benz14}
{Benz} W.,  {Ida} S.,  {Alibert} Y.,  {Lin} D.,   {Mordasini} C.,  2014, in
  {Beuther} H.,  {Klessen} R.~S.,  {Dullemond} C.~P.,   {Henning} T.,  eds,
  Protostars and Planets VI. p.~691 (\mn@eprint {arXiv} {1402.7086}),
  \mn@doi{10.2458/azu\_uapress\_9780816531240-ch030}

\bibitem[\protect\citeauthoryear{{Birnstiel}, {Klahr}  \&
  {Ercolano}}{{Birnstiel} et~al.}{2012}]{Birnstiel12}
{Birnstiel} T.,  {Klahr} H.,   {Ercolano} B.,  2012, \mn@doi [\aap]
  {10.1051/0004-6361/201118136}, \href
  {https://ui.adsabs.harvard.edu/abs/2012A&A...539A.148B} {539, A148}

\bibitem[\protect\citeauthoryear{{Bitsch}, {Johansen}, {Lambrechts}  \&
  {Morbidelli}}{{Bitsch} et~al.}{2015}]{Bitsch15a}
{Bitsch} B.,  {Johansen} A.,  {Lambrechts} M.,   {Morbidelli} A.,  2015,
  \mn@doi [\aap] {10.1051/0004-6361/201424964}, \href
  {https://ui.adsabs.harvard.edu/abs/2015A&A...575A..28B} {575, A28}

\bibitem[\protect\citeauthoryear{{Bodenheimer} \& {Pollack}}{{Bodenheimer} \&
  {Pollack}}{1986}]{Bodenheimer86}
{Bodenheimer} P.,  {Pollack} J.~B.,  1986, \mn@doi [\icarus]
  {10.1016/0019-1035(86)90122-3}, \href
  {https://ui.adsabs.harvard.edu/abs/1986Icar...67..391B} {67, 391}

\bibitem[\protect\citeauthoryear{{Boyden} \& {Eisner}}{{Boyden} \&
  {Eisner}}{2020}]{Boyden20}
{Boyden} R.~D.,  {Eisner} J.~A.,  2020, \mn@doi [\apj]
  {10.3847/1538-4357/ab86b7}, \href
  {https://ui.adsabs.harvard.edu/abs/2020ApJ...894...74B} {894, 74}

\bibitem[\protect\citeauthoryear{{Brucalassi} et~al.,}{{Brucalassi}
  et~al.}{2016}]{Brucalassi16}
{Brucalassi} A.,  et~al., 2016, \mn@doi [\aap] {10.1051/0004-6361/201527561},
  \href {https://ui.adsabs.harvard.edu/abs/2016A&A...592L...1B} {592, L1}

\bibitem[\protect\citeauthoryear{{Clarke}}{{Clarke}}{2007}]{Cla07}
{Clarke} C.~J.,  2007, \mn@doi [\mnras] {10.1111/j.1365-2966.2007.11547.x},
  376, 1350

\bibitem[\protect\citeauthoryear{{Clarke}, {Gendrin}  \& {Sotomayor}}{{Clarke}
  et~al.}{2001}]{Cla01}
{Clarke} C.~J.,  {Gendrin} A.,   {Sotomayor} M.,  2001, \mn@doi [\mnras]
  {10.1046/j.1365-8711.2001.04891.x}, \href
  {https://ui.adsabs.harvard.edu/abs/2001MNRAS.328..485C} {328, 485}

\bibitem[\protect\citeauthoryear{{Coleman}}{{Coleman}}{2021}]{Coleman21}
{Coleman} G. A.~L.,  2021, \mn@doi [\mnras] {10.1093/mnras/stab1904}, \href
  {https://ui.adsabs.harvard.edu/abs/2021MNRAS.506.3596C} {506, 3596}

\bibitem[\protect\citeauthoryear{{Coleman} \& {Haworth}}{{Coleman} \&
  {Haworth}}{2022}]{Coleman22}
{Coleman} G. A.~L.,  {Haworth} T.~J.,  2022, arXiv e-prints, \href
  {https://ui.adsabs.harvard.edu/abs/2022arXiv220402303C} {p. arXiv:2204.02303}

\bibitem[\protect\citeauthoryear{{Coleman} \& {Nelson}}{{Coleman} \&
  {Nelson}}{2014}]{Coleman14}
{Coleman} G. A.~L.,  {Nelson} R.~P.,  2014, \mn@doi [\mnras]
  {10.1093/mnras/stu1715}, \href
  {https://ui.adsabs.harvard.edu/abs/2014MNRAS.445..479C} {445, 479}

\bibitem[\protect\citeauthoryear{{Coleman} \& {Nelson}}{{Coleman} \&
  {Nelson}}{2016}]{Coleman16}
{Coleman} G. A.~L.,  {Nelson} R.~P.,  2016, \mn@doi [\mnras]
  {10.1093/mnras/stw1177}, \href
  {https://ui.adsabs.harvard.edu/abs/2016MNRAS.460.2779C} {460, 2779}

\bibitem[\protect\citeauthoryear{{Concha-Ram{\'\i}rez}, {Wilhelm}, {Portegies
  Zwart}  \& {Haworth}}{{Concha-Ram{\'\i}rez} et~al.}{2019}]{Concha-Ramirez19}
{Concha-Ram{\'\i}rez} F.,  {Wilhelm} M. J.~C.,  {Portegies Zwart} S.,
  {Haworth} T.~J.,  2019, \mn@doi [\mnras] {10.1093/mnras/stz2973}, \href
  {https://ui.adsabs.harvard.edu/abs/2019MNRAS.490.5678C} {490, 5678}

\bibitem[\protect\citeauthoryear{{Crida}, {Morbidelli}  \& {Masset}}{{Crida}
  et~al.}{2006}]{Crida06}
{Crida} A.,  {Morbidelli} A.,   {Masset} F.,  2006, \mn@doi [\icarus]
  {10.1016/j.icarus.2005.10.007}, \href
  {https://ui.adsabs.harvard.edu/abs/2006Icar..181..587C} {181, 587}

\bibitem[\protect\citeauthoryear{{D'Angelo}, {Henning}  \& {Kley}}{{D'Angelo}
  et~al.}{2002}]{Dangelo02}
{D'Angelo} G.,  {Henning} T.,   {Kley} W.,  2002, \mn@doi [\aap]
  {10.1051/0004-6361:20020173}, \href
  {https://ui.adsabs.harvard.edu/abs/2002A&A...385..647D} {385, 647}

\bibitem[\protect\citeauthoryear{{Drazkowska} et~al.,}{{Drazkowska}
  et~al.}{2022}]{Drazkowska22}
{Drazkowska} J.,  et~al., 2022, arXiv e-prints, \href
  {https://ui.adsabs.harvard.edu/abs/2022arXiv220309759D} {p. arXiv:2203.09759}

\bibitem[\protect\citeauthoryear{{Dr{\k{a}}{\.z}kowska} \&
  {Dullemond}}{{Dr{\k{a}}{\.z}kowska} \& {Dullemond}}{2014}]{Drakowska14}
{Dr{\k{a}}{\.z}kowska} J.,  {Dullemond} C.~P.,  2014, \mn@doi [\aap]
  {10.1051/0004-6361/201424809}, \href
  {https://ui.adsabs.harvard.edu/abs/2014A&A...572A..78D} {572, A78}

\bibitem[\protect\citeauthoryear{{Dr{\k{a}}{\.z}kowska}, {Alibert}  \&
  {Moore}}{{Dr{\k{a}}{\.z}kowska} et~al.}{2016}]{Drazkowska16}
{Dr{\k{a}}{\.z}kowska} J.,  {Alibert} Y.,   {Moore} B.,  2016, \mn@doi [\aap]
  {10.1051/0004-6361/201628983}, \href
  {https://ui.adsabs.harvard.edu/abs/2016A&A...594A.105D} {594, A105}

\bibitem[\protect\citeauthoryear{{Dr{\k{a}}{\.z}kowska}, {Stammler}  \&
  {Birnstiel}}{{Dr{\k{a}}{\.z}kowska} et~al.}{2021}]{Drazkowska21}
{Dr{\k{a}}{\.z}kowska} J.,  {Stammler} S.~M.,   {Birnstiel} T.,  2021, \mn@doi
  [\aap] {10.1051/0004-6361/202039925}, \href
  {https://ui.adsabs.harvard.edu/abs/2021A&A...647A..15D} {647, A15}

\bibitem[\protect\citeauthoryear{{Eisner} et~al.,}{{Eisner}
  et~al.}{2018}]{Eis18}
{Eisner} J.~A.,  et~al., 2018, \mn@doi [\apj] {10.3847/1538-4357/aac3e2}, 860,
  77

\bibitem[\protect\citeauthoryear{{Emsenhuber}, {Mordasini}, {Burn}, {Alibert},
  {Benz}  \& {Asphaug}}{{Emsenhuber} et~al.}{2021}]{Emsenhuber21}
{Emsenhuber} A.,  {Mordasini} C.,  {Burn} R.,  {Alibert} Y.,  {Benz} W.,
  {Asphaug} E.,  2021, \mn@doi [\aap] {10.1051/0004-6361/202038553}, \href
  {https://ui.adsabs.harvard.edu/abs/2021A&A...656A..69E} {656, A69}

\bibitem[\protect\citeauthoryear{{Ercolano} \& {Pascucci}}{{Ercolano} \&
  {Pascucci}}{2017}]{Ercolano17}
{Ercolano} B.,  {Pascucci} I.,  2017, \mn@doi [Royal Society Open Science]
  {10.1098/rsos.170114}, \href
  {https://ui.adsabs.harvard.edu/abs/2017RSOS....470114E} {4, 170114}

\bibitem[\protect\citeauthoryear{{Ercolano} \& {Rosotti}}{{Ercolano} \&
  {Rosotti}}{2015}]{Ercolano15}
{Ercolano} B.,  {Rosotti} G.,  2015, \mn@doi [\mnras] {10.1093/mnras/stv833},
  \href {https://ui.adsabs.harvard.edu/abs/2015MNRAS.450.3008E} {450, 3008}

\bibitem[\protect\citeauthoryear{{Facchini}, {Clarke}  \& {Bisbas}}{{Facchini}
  et~al.}{2016}]{Fac16}
{Facchini} S.,  {Clarke} C.~J.,   {Bisbas} T.~G.,  2016, \mn@doi [\mnras]
  {10.1093/mnras/stw240}, 457, 3593

\bibitem[\protect\citeauthoryear{{Fang} et~al.,}{{Fang} et~al.}{2012}]{Fang12}
{Fang} M.,  et~al., 2012, \mn@doi [\aap] {10.1051/0004-6361/201015914}, \href
  {https://ui.adsabs.harvard.edu/abs/2012A&A...539A.119F} {539, A119}

\bibitem[\protect\citeauthoryear{{Fatuzzo} \& {Adams}}{{Fatuzzo} \&
  {Adams}}{2008}]{Fat08}
{Fatuzzo} M.,  {Adams} F.~C.,  2008, \mn@doi [\apj] {10.1086/527469}, 675, 1361

\bibitem[\protect\citeauthoryear{{Fernandes}, {Mulders}, {Pascucci},
  {Mordasini}  \& {Emsenhuber}}{{Fernandes} et~al.}{2019}]{Fernandes19}
{Fernandes} R.~B.,  {Mulders} G.~D.,  {Pascucci} I.,  {Mordasini} C.,
  {Emsenhuber} A.,  2019, \mn@doi [\apj] {10.3847/1538-4357/ab0300}, \href
  {https://ui.adsabs.harvard.edu/abs/2019ApJ...874...81F} {874, 81}

\bibitem[\protect\citeauthoryear{{Getman}, {Feigelson}  \& {Kuhn}}{{Getman}
  et~al.}{2014}]{Get14}
{Getman} K.~V.,  {Feigelson} E.~D.,   {Kuhn} M.~A.,  2014, \mn@doi [\apj]
  {10.1088/0004-637X/787/2/109}, 787, 109

\bibitem[\protect\citeauthoryear{{Gilliland} et~al.,}{{Gilliland}
  et~al.}{2000}]{Gil00}
{Gilliland} R.~L.,  et~al., 2000, \mn@doi [\apjl] {10.1086/317334}, 545, L47

\bibitem[\protect\citeauthoryear{{Goldreich} \& {Tremaine}}{{Goldreich} \&
  {Tremaine}}{1980}]{Goldreich80}
{Goldreich} P.,  {Tremaine} S.,  1980, \mn@doi [\apj] {10.1086/158356}, \href
  {https://ui.adsabs.harvard.edu/abs/1980ApJ...241..425G} {241, 425}

\bibitem[\protect\citeauthoryear{{Guarcello} et~al.,}{{Guarcello}
  et~al.}{2016}]{Gua16}
{Guarcello} M.~G.,  et~al., 2016, preprint (\mn@eprint {arXiv} {1605.01773})

\bibitem[\protect\citeauthoryear{{Guilera}, {Cuello}, {Montesinos}, {Miller
  Bertolami}, {Ronco}, {Cuadra}  \& {Masset}}{{Guilera}
  et~al.}{2019}]{Guilera19}
{Guilera} O.~M.,  {Cuello} N.,  {Montesinos} M.,  {Miller Bertolami} M.~M.,
  {Ronco} M.~P.,  {Cuadra} J.,   {Masset} F.~S.,  2019, \mn@doi [\mnras]
  {10.1093/mnras/stz1158}, \href
  {https://ui.adsabs.harvard.edu/abs/2019MNRAS.486.5690G} {486, 5690}

\bibitem[\protect\citeauthoryear{{Guilera}, {Miller Bertolami}, {Masset},
  {Cuadra}, {Venturini}  \& {Ronco}}{{Guilera} et~al.}{2021}]{Guilera21}
{Guilera} O.~M.,  {Miller Bertolami} M.~M.,  {Masset} F.,  {Cuadra} J.,
  {Venturini} J.,   {Ronco} M.~P.,  2021, \mn@doi [\mnras]
  {10.1093/mnras/stab2371}, \href
  {https://ui.adsabs.harvard.edu/abs/2021MNRAS.507.3638G} {507, 3638}

\bibitem[\protect\citeauthoryear{{Habing}}{{Habing}}{1968}]{Hab68}
{Habing} H.~J.,  1968, \bain, 19, 421

\bibitem[\protect\citeauthoryear{{Haisch}, {Lada}, {Pi{\~n}a}, {Telesco}  \&
  {Lada}}{{Haisch} et~al.}{2001}]{Hai01}
{Haisch} Jr. K.~E.,  {Lada} E.~A.,  {Pi{\~n}a} R.~K.,  {Telesco} C.~M.,
  {Lada} C.~J.,  2001, \mn@doi [\aj] {10.1086/319397}, 121, 1512

\bibitem[\protect\citeauthoryear{Harris et~al.,}{Harris
  et~al.}{2020}]{Harris20}
Harris C.~R.,  et~al., 2020, \mn@doi [Nature] {10.1038/s41586-020-2649-2}, 585,
  357

\bibitem[\protect\citeauthoryear{{Hartmann}, {Calvet}, {Gullbring}  \&
  {D'Alessio}}{{Hartmann} et~al.}{1998}]{Hartmann98}
{Hartmann} L.,  {Calvet} N.,  {Gullbring} E.,   {D'Alessio} P.,  1998, \mn@doi
  [\apj] {10.1086/305277}, \href
  {https://ui.adsabs.harvard.edu/abs/1998ApJ...495..385H} {495, 385}

\bibitem[\protect\citeauthoryear{{Hasegawa} et~al.,}{{Hasegawa}
  et~al.}{2022}]{Hasegawa21}
{Hasegawa} Y.,  et~al., 2022, \mn@doi [\apjl] {10.3847/2041-8213/ac50aa}, \href
  {https://ui.adsabs.harvard.edu/abs/2022ApJ...926L..23H} {926, L23}

\bibitem[\protect\citeauthoryear{{Haworth}, {Facchini}, {Clarke}  \&
  {Mohanty}}{{Haworth} et~al.}{2018a}]{Haw18}
{Haworth} T.~J.,  {Facchini} S.,  {Clarke} C.~J.,   {Mohanty} S.,  2018a,
  \mn@doi [\mnras] {10.1093/mnras/sty168}

\bibitem[\protect\citeauthoryear{{Haworth}, {Clarke}, {Rahman}, {Winter}  \&
  {Facchini}}{{Haworth} et~al.}{2018b}]{Haw18b}
{Haworth} T.~J.,  {Clarke} C.~J.,  {Rahman} W.,  {Winter} A.~J.,   {Facchini}
  S.,  2018b, \mn@doi [\mnras] {10.1093/mnras/sty2323}, 481, 452

\bibitem[\protect\citeauthoryear{{Haworth}, {Kim}, {Winter}, {Hines}, {Clarke},
  {Sellek}, {Ballabio}  \& {Stapelfeldt}}{{Haworth} et~al.}{2021}]{Haworth21}
{Haworth} T.~J.,  {Kim} J.~S.,  {Winter} A.~J.,  {Hines} D.~C.,  {Clarke}
  C.~J.,  {Sellek} A.~D.,  {Ballabio} G.,   {Stapelfeldt} K.~R.,  2021, \mn@doi
  [\mnras] {10.1093/mnras/staa3918}, \href
  {https://ui.adsabs.harvard.edu/abs/2021MNRAS.501.3502H} {501, 3502}

\bibitem[\protect\citeauthoryear{{Hayashi}}{{Hayashi}}{1981}]{Hayashi81}
{Hayashi} C.,  1981, \mn@doi [Progress of Theoretical Physics Supplement]
  {10.1143/PTPS.70.35}, \href
  {https://ui.adsabs.harvard.edu/abs/1981PThPS..70...35H} {70, 35}

\bibitem[\protect\citeauthoryear{{Henning} \& {Stognienko}}{{Henning} \&
  {Stognienko}}{1996}]{Henning96}
{Henning} T.,  {Stognienko} R.,  1996, \aap, \href
  {https://ui.adsabs.harvard.edu/abs/1996A&A...311..291H} {311, 291}

\bibitem[\protect\citeauthoryear{{Hillenbrand} \& {Hartmann}}{{Hillenbrand} \&
  {Hartmann}}{1998}]{Hil98}
{Hillenbrand} L.~A.,  {Hartmann} L.~W.,  1998, \mn@doi [\apj] {10.1086/305076},
  492, 540

\bibitem[\protect\citeauthoryear{{Hollenbach}, {Johnstone}, {Lizano}  \&
  {Shu}}{{Hollenbach} et~al.}{1994}]{Hol94}
{Hollenbach} D.,  {Johnstone} D.,  {Lizano} S.,   {Shu} F.,  1994, \mn@doi
  [\apj] {10.1086/174276}, 428, 654

\bibitem[\protect\citeauthoryear{Hunter}{Hunter}{2007}]{Hunter07}
Hunter J.~D.,  2007, \mn@doi [Computing in Science \& Engineering]
  {10.1109/MCSE.2007.55}, 9, 90

\bibitem[\protect\citeauthoryear{{Ida} \& {Lin}}{{Ida} \&
  {Lin}}{2004a}]{Ida04a}
{Ida} S.,  {Lin} D.~N.~C.,  2004a, \mn@doi [\apj] {10.1086/381724}, \href
  {https://ui.adsabs.harvard.edu/abs/2004ApJ...604..388I} {604, 388}

\bibitem[\protect\citeauthoryear{{Ida} \& {Lin}}{{Ida} \&
  {Lin}}{2004b}]{Ida04b}
{Ida} S.,  {Lin} D.~N.~C.,  2004b, \mn@doi [\apj] {10.1086/424830}, \href
  {https://ui.adsabs.harvard.edu/abs/2004ApJ...616..567I} {616, 567}

\bibitem[\protect\citeauthoryear{{Ida}, {Tanaka}, {Johansen}, {Kanagawa}  \&
  {Tanigawa}}{{Ida} et~al.}{2018}]{Ida18}
{Ida} S.,  {Tanaka} H.,  {Johansen} A.,  {Kanagawa} K.~D.,   {Tanigawa} T.,
  2018, \mn@doi [\apj] {10.3847/1538-4357/aad69c}, \href
  {https://ui.adsabs.harvard.edu/abs/2018ApJ...864...77I} {864, 77}

\bibitem[\protect\citeauthoryear{{Ikoma}, {Nakazawa}  \& {Emori}}{{Ikoma}
  et~al.}{2000}]{Ikoma00}
{Ikoma} M.,  {Nakazawa} K.,   {Emori} H.,  2000, \mn@doi [\apj]
  {10.1086/309050}, \href
  {https://ui.adsabs.harvard.edu/abs/2000ApJ...537.1013I} {537, 1013}

\bibitem[\protect\citeauthoryear{{Jennings}, {Ercolano}  \&
  {Rosotti}}{{Jennings} et~al.}{2018}]{Jennings18}
{Jennings} J.,  {Ercolano} B.,   {Rosotti} G.~P.,  2018, \mn@doi [\mnras]
  {10.1093/mnras/sty964}, \href
  {https://ui.adsabs.harvard.edu/abs/2018MNRAS.477.4131J} {477, 4131}

\bibitem[\protect\citeauthoryear{{Johnson}, {Aller}, {Howard}  \&
  {Crepp}}{{Johnson} et~al.}{2010}]{Johnson10}
{Johnson} J.~A.,  {Aller} K.~M.,  {Howard} A.~W.,   {Crepp} J.~R.,  2010,
  \mn@doi [\pasp] {10.1086/655775}, \href
  {https://ui.adsabs.harvard.edu/abs/2010PASP..122..905J} {122, 905}

\bibitem[\protect\citeauthoryear{{Johnstone}, {Hollenbach}  \&
  {Bally}}{{Johnstone} et~al.}{1998}]{Johnstone98}
{Johnstone} D.,  {Hollenbach} D.,   {Bally} J.,  1998, \mn@doi [\apj]
  {10.1086/305658}, \href
  {https://ui.adsabs.harvard.edu/abs/1998ApJ...499..758J} {499, 758}

\bibitem[\protect\citeauthoryear{{Kim}, {Clarke}, {Fang}  \& {Facchini}}{{Kim}
  et~al.}{2016}]{Kim16}
{Kim} J.~S.,  {Clarke} C.~J.,  {Fang} M.,   {Facchini} S.,  2016, \mn@doi
  [\apjl] {10.3847/2041-8205/826/1/L15}, 826, L15

\bibitem[\protect\citeauthoryear{{Kimmig}, {Dullemond}  \& {Kley}}{{Kimmig}
  et~al.}{2020}]{Kimmig20}
{Kimmig} C.~N.,  {Dullemond} C.~P.,   {Kley} W.,  2020, \mn@doi [\aap]
  {10.1051/0004-6361/201936412}, \href
  {https://ui.adsabs.harvard.edu/abs/2020A&A...633A...4K} {633, A4}

\bibitem[\protect\citeauthoryear{{Kruijssen}, {Longmore}  \&
  {Chevance}}{{Kruijssen} et~al.}{2020}]{Kruijssen20}
{Kruijssen} J.~M.~D.,  {Longmore} S.~N.,   {Chevance} M.,  2020, \mn@doi
  [\apjl] {10.3847/2041-8213/abccc3}, \href
  {https://ui.adsabs.harvard.edu/abs/2020ApJ...905L..18K} {905, L18}

\bibitem[\protect\citeauthoryear{{Lada} \& {Lada}}{{Lada} \&
  {Lada}}{2003}]{Lada03}
{Lada} C.~J.,  {Lada} E.~A.,  2003, \mn@doi [\araa]
  {10.1146/annurev.astro.41.011802.094844}, \href
  {https://ui.adsabs.harvard.edu/abs/2003ARA&A..41...57L} {41, 57}

\bibitem[\protect\citeauthoryear{{Lambrechts} \& {Johansen}}{{Lambrechts} \&
  {Johansen}}{2012}]{Lambrechts12}
{Lambrechts} M.,  {Johansen} A.,  2012, \mn@doi [\aap]
  {10.1051/0004-6361/201219127}, \href
  {https://ui.adsabs.harvard.edu/abs/2012A&A...544A..32L} {544, A32}

\bibitem[\protect\citeauthoryear{{Lambrechts} \& {Johansen}}{{Lambrechts} \&
  {Johansen}}{2014}]{Lambrechts14}
{Lambrechts} M.,  {Johansen} A.,  2014, \mn@doi [\aap]
  {10.1051/0004-6361/201424343}, \href
  {https://ui.adsabs.harvard.edu/abs/2014A&A...572A.107L} {572, A107}

\bibitem[\protect\citeauthoryear{{Larsen}}{{Larsen}}{2002}]{Larsen02}
{Larsen} S.~S.,  2002, \mn@doi [\aj] {10.1086/342381}, \href
  {https://ui.adsabs.harvard.edu/abs/2002AJ....124.1393L} {124, 1393}

\bibitem[\protect\citeauthoryear{{Liffman}}{{Liffman}}{2003}]{Liffman03}
{Liffman} K.,  2003, \mn@doi [\pasa] {10.1071/AS03019}, \href
  {https://ui.adsabs.harvard.edu/abs/2003PASA...20..337L} {20, 337}

\bibitem[\protect\citeauthoryear{{Lin} \& {Papaloizou}}{{Lin} \&
  {Papaloizou}}{1986}]{Lin86}
{Lin} D.~N.~C.,  {Papaloizou} J.,  1986, \mn@doi [\apj] {10.1086/164653}, \href
  {https://ui.adsabs.harvard.edu/abs/1986ApJ...309..846L} {309, 846}

\bibitem[\protect\citeauthoryear{{Lubow}, {Seibert}  \& {Artymowicz}}{{Lubow}
  et~al.}{1999}]{Lubow99}
{Lubow} S.~H.,  {Seibert} M.,   {Artymowicz} P.,  1999, \mn@doi [\apj]
  {10.1086/308045}, \href
  {https://ui.adsabs.harvard.edu/abs/1999ApJ...526.1001L} {526, 1001}

\bibitem[\protect\citeauthoryear{{Lynden-Bell} \& {Pringle}}{{Lynden-Bell} \&
  {Pringle}}{1974}]{Lynden-Bell74}
{Lynden-Bell} D.,  {Pringle} J.~E.,  1974, \mn@doi [\mnras]
  {10.1093/mnras/168.3.603}, \href
  {https://ui.adsabs.harvard.edu/abs/1974MNRAS.168..603L} {168, 603}

\bibitem[\protect\citeauthoryear{{Manara} et~al.,}{{Manara}
  et~al.}{2016}]{Manara16}
{Manara} C.~F.,  et~al., 2016, \mn@doi [\aap] {10.1051/0004-6361/201628549},
  \href {https://ui.adsabs.harvard.edu/abs/2016A&A...591L...3M} {591, L3}

\bibitem[\protect\citeauthoryear{{Manara} et~al.,}{{Manara}
  et~al.}{2017}]{Manara17}
{Manara} C.~F.,  et~al., 2017, \mn@doi [\aap] {10.1051/0004-6361/201630147},
  \href {https://ui.adsabs.harvard.edu/abs/2017A&A...604A.127M} {604, A127}

\bibitem[\protect\citeauthoryear{{Matsuyama}, {Johnstone}  \&
  {Murray}}{{Matsuyama} et~al.}{2003}]{Matsuyama03}
{Matsuyama} I.,  {Johnstone} D.,   {Murray} N.,  2003, \mn@doi [\apjl]
  {10.1086/374406}, \href
  {https://ui.adsabs.harvard.edu/abs/2003ApJ...585L.143M} {585, L143}

\bibitem[\protect\citeauthoryear{{Mayor} et~al.,}{{Mayor}
  et~al.}{2011}]{Mayor11}
{Mayor} M.,  et~al., 2011, arXiv e-prints, \href
  {https://ui.adsabs.harvard.edu/abs/2011arXiv1109.2497M} {p. arXiv:1109.2497}

\bibitem[\protect\citeauthoryear{{Miller} \& {Scalo}}{{Miller} \&
  {Scalo}}{1978}]{Miller78}
{Miller} G.~E.,  {Scalo} J.~M.,  1978, \mn@doi [\pasp] {10.1086/130373}, \href
  {https://ui.adsabs.harvard.edu/abs/1978PASP...90..506M} {90, 506}

\bibitem[\protect\citeauthoryear{{Miotello}, {Robberto}, {Potenza}  \&
  {Ricci}}{{Miotello} et~al.}{2012}]{Miotello12}
{Miotello} A.,  {Robberto} M.,  {Potenza} M. A.~C.,   {Ricci} L.,  2012,
  \mn@doi [\apj] {10.1088/0004-637X/757/1/78}, \href
  {https://ui.adsabs.harvard.edu/abs/2012ApJ...757...78M} {757, 78}

\bibitem[\protect\citeauthoryear{{Monsch}, {Picogna}, {Ercolano}  \&
  {Kley}}{{Monsch} et~al.}{2021a}]{Monsch21}
{Monsch} K.,  {Picogna} G.,  {Ercolano} B.,   {Kley} W.,  2021a, \mn@doi [\aap]
  {10.1051/0004-6361/202039658}, \href
  {https://ui.adsabs.harvard.edu/abs/2021A&A...646A.169M} {646, A169}

\bibitem[\protect\citeauthoryear{{Monsch}, {Picogna}, {Ercolano}  \&
  {Preibisch}}{{Monsch} et~al.}{2021b}]{Monsch21b}
{Monsch} K.,  {Picogna} G.,  {Ercolano} B.,   {Preibisch} T.,  2021b, \mn@doi
  [\aap] {10.1051/0004-6361/202140647}, \href
  {https://ui.adsabs.harvard.edu/abs/2021A&A...650A.199M} {650, A199}

\bibitem[\protect\citeauthoryear{{Mordasini}}{{Mordasini}}{2018}]{Mordasini18}
{Mordasini} C.,  2018, in {Deeg} H.~J.,  {Belmonte} J.~A.,  eds, , Handbook of
  Exoplanets.
p.~143, \mn@doi{10.1007/978-3-319-55333-7\_143}

\bibitem[\protect\citeauthoryear{{Nakatani}, {Kobayashi}, {Kuiper}, {Nomura}
  \& {Aikawa}}{{Nakatani} et~al.}{2021}]{Nakatani21}
{Nakatani} R.,  {Kobayashi} H.,  {Kuiper} R.,  {Nomura} H.,   {Aikawa} Y.,
  2021, \mn@doi [\apj] {10.3847/1538-4357/ac0137}, \href
  {https://ui.adsabs.harvard.edu/abs/2021ApJ...915...90N} {915, 90}

\bibitem[\protect\citeauthoryear{{Ndugu}, {Bitsch}  \& {Jurua}}{{Ndugu}
  et~al.}{2018}]{Ndugu18}
{Ndugu} N.,  {Bitsch} B.,   {Jurua} E.,  2018, \mn@doi [\mnras]
  {10.1093/mnras/stx2815}, \href
  {https://ui.adsabs.harvard.edu/abs/2018MNRAS.474..886N} {474, 886}

\bibitem[\protect\citeauthoryear{{Nelson} \& {Papaloizou}}{{Nelson} \&
  {Papaloizou}}{2004}]{Nelson04}
{Nelson} R.~P.,  {Papaloizou} J. C.~B.,  2004, \mn@doi [\mnras]
  {10.1111/j.1365-2966.2004.07406.x}, \href
  {https://ui.adsabs.harvard.edu/abs/2004MNRAS.350..849N} {350, 849}

\bibitem[\protect\citeauthoryear{{O'Dell} \& {Wen}}{{O'Dell} \&
  {Wen}}{1994}]{Ode94}
{O'Dell} C.~R.,  {Wen} Z.,  1994, \mn@doi [\apj] {10.1086/174892}, 436, 194

\bibitem[\protect\citeauthoryear{{Owen}, {Ercolano}, {Clarke}  \&
  {Alexander}}{{Owen} et~al.}{2010}]{Owen10}
{Owen} J.~E.,  {Ercolano} B.,  {Clarke} C.~J.,   {Alexander} R.~D.,  2010,
  \mn@doi [\mnras] {10.1111/j.1365-2966.2009.15771.x}, \href
  {https://ui.adsabs.harvard.edu/abs/2010MNRAS.401.1415O} {401, 1415}

\bibitem[\protect\citeauthoryear{{Owen}, {Clarke}  \& {Ercolano}}{{Owen}
  et~al.}{2012}]{Owen12}
{Owen} J.~E.,  {Clarke} C.~J.,   {Ercolano} B.,  2012, \mn@doi [\mnras]
  {10.1111/j.1365-2966.2011.20337.x}, \href
  {https://ui.adsabs.harvard.edu/abs/2012MNRAS.422.1880O} {422, 1880}

\bibitem[\protect\citeauthoryear{{Paardekooper}, {Baruteau}  \&
  {Kley}}{{Paardekooper} et~al.}{2011}]{Paardekooper11}
{Paardekooper} S.~J.,  {Baruteau} C.,   {Kley} W.,  2011, \mn@doi [\mnras]
  {10.1111/j.1365-2966.2010.17442.x}, \href
  {https://ui.adsabs.harvard.edu/abs/2011MNRAS.410..293P} {410, 293}

\bibitem[\protect\citeauthoryear{{Pascucci}, {Ricci}, {Gorti}, {Hollenbach},
  {Hendler}, {Brooks}  \& {Contreras}}{{Pascucci} et~al.}{2014}]{Pascucci14}
{Pascucci} I.,  {Ricci} L.,  {Gorti} U.,  {Hollenbach} D.,  {Hendler} N.~P.,
  {Brooks} K.~J.,   {Contreras} Y.,  2014, \mn@doi [\apj]
  {10.1088/0004-637X/795/1/1}, \href
  {https://ui.adsabs.harvard.edu/abs/2014ApJ...795....1P} {795, 1}

\bibitem[\protect\citeauthoryear{{Picogna}, {Ercolano}  \&
  {Espaillat}}{{Picogna} et~al.}{2021}]{Picogna21}
{Picogna} G.,  {Ercolano} B.,   {Espaillat} C.~C.,  2021, \mn@doi [\mnras]
  {10.1093/mnras/stab2883}, \href
  {https://ui.adsabs.harvard.edu/abs/2021MNRAS.508.3611P} {508, 3611}

\bibitem[\protect\citeauthoryear{{Pinilla}, {Benisty}  \&
  {Birnstiel}}{{Pinilla} et~al.}{2012}]{Pinilla12}
{Pinilla} P.,  {Benisty} M.,   {Birnstiel} T.,  2012, \mn@doi [\aap]
  {10.1051/0004-6361/201219315}, \href
  {https://ui.adsabs.harvard.edu/abs/2012A&A...545A..81P} {545, A81}

\bibitem[\protect\citeauthoryear{{Pinte}, {Dent}, {M{\'e}nard}, {Hales},
  {Hill}, {Cortes}  \& {de Gregorio-Monsalvo}}{{Pinte} et~al.}{2016}]{Pinte16}
{Pinte} C.,  {Dent} W.~R.~F.,  {M{\'e}nard} F.,  {Hales} A.,  {Hill} T.,
  {Cortes} P.,   {de Gregorio-Monsalvo} I.,  2016, \mn@doi [\apj]
  {10.3847/0004-637X/816/1/25}, \href
  {https://ui.adsabs.harvard.edu/abs/2016ApJ...816...25P} {816, 25}

\bibitem[\protect\citeauthoryear{{Pollack}, {Hubickyj}, {Bodenheimer},
  {Lissauer}, {Podolak}  \& {Greenzweig}}{{Pollack} et~al.}{1996}]{Pollack96}
{Pollack} J.~B.,  {Hubickyj} O.,  {Bodenheimer} P.,  {Lissauer} J.~J.,
  {Podolak} M.,   {Greenzweig} Y.,  1996, \mn@doi [\icarus]
  {10.1006/icar.1996.0190}, \href
  {https://ui.adsabs.harvard.edu/abs/1996Icar..124...62P} {124, 62}

\bibitem[\protect\citeauthoryear{{Qiao}, {Haworth}, {Sellek}  \& {Ali}}{{Qiao}
  et~al.}{2022}]{Qiao22}
{Qiao} L.,  {Haworth} T.~J.,  {Sellek} A.~D.,   {Ali} A.~A.,  2022, \mn@doi
  [\mnras] {10.1093/mnras/stac684}, \href
  {https://ui.adsabs.harvard.edu/abs/2022MNRAS.512.3788Q} {512, 3788}

\bibitem[\protect\citeauthoryear{{Rafikov}}{{Rafikov}}{2004}]{Rafikov04}
{Rafikov} R.~R.,  2004, \mn@doi [\aj] {10.1086/423216}, \href
  {https://ui.adsabs.harvard.edu/abs/2004AJ....128.1348R} {128, 1348}

\bibitem[\protect\citeauthoryear{{Reffert}, {Bergmann}, {Quirrenbach},
  {Trifonov}  \& {K{\"u}nstler}}{{Reffert} et~al.}{2015}]{Reffert15}
{Reffert} S.,  {Bergmann} C.,  {Quirrenbach} A.,  {Trifonov} T.,
  {K{\"u}nstler} A.,  2015, \mn@doi [\aap] {10.1051/0004-6361/201322360}, \href
  {https://ui.adsabs.harvard.edu/abs/2015A&A...574A.116R} {574, A116}

\bibitem[\protect\citeauthoryear{{Rodenkirch}, {Klahr}, {Fendt}  \&
  {Dullemond}}{{Rodenkirch} et~al.}{2020}]{Rodenkirch20}
{Rodenkirch} P.~J.,  {Klahr} H.,  {Fendt} C.,   {Dullemond} C.~P.,  2020,
  \mn@doi [\aap] {10.1051/0004-6361/201834945}, \href
  {https://ui.adsabs.harvard.edu/abs/2020A&A...633A..21R} {633, A21}

\bibitem[\protect\citeauthoryear{{Schlaufman}}{{Schlaufman}}{2018}]{Schlaufman18}
{Schlaufman} K.~C.,  2018, \mn@doi [\apj] {10.3847/1538-4357/aa961c}, \href
  {https://ui.adsabs.harvard.edu/abs/2018ApJ...853...37S} {853, 37}

\bibitem[\protect\citeauthoryear{{Sellek}, {Booth}  \& {Clarke}}{{Sellek}
  et~al.}{2020}]{Sellek20}
{Sellek} A.~D.,  {Booth} R.~A.,   {Clarke} C.~J.,  2020, \mn@doi [\mnras]
  {10.1093/mnras/stz3528}, \href
  {https://ui.adsabs.harvard.edu/abs/2020MNRAS.492.1279S} {492, 1279}

\bibitem[\protect\citeauthoryear{{Stolte}, {Brandner}, {Brandl}, {Zinnecker}
  \& {Grebel}}{{Stolte} et~al.}{2004}]{Stolte04}
{Stolte} A.,  {Brandner} W.,  {Brandl} B.,  {Zinnecker} H.,   {Grebel} E.~K.,
  2004, \mn@doi [\aj] {10.1086/422705}, \href
  {https://ui.adsabs.harvard.edu/abs/2004AJ....128..765S} {128, 765}

\bibitem[\protect\citeauthoryear{{St{\"o}rzer} \& {Hollenbach}}{{St{\"o}rzer}
  \& {Hollenbach}}{1999}]{Sto99}
{St{\"o}rzer} H.,  {Hollenbach} D.,  1999, \mn@doi [\apj] {10.1086/307055},
  515, 669

\bibitem[\protect\citeauthoryear{{Suzuki} \& {Inutsuka}}{{Suzuki} \&
  {Inutsuka}}{2009}]{Suzuki09}
{Suzuki} T.~K.,  {Inutsuka} S.-i.,  2009, \mn@doi [\apjl]
  {10.1088/0004-637X/691/1/L49}, \href
  {https://ui.adsabs.harvard.edu/abs/2009ApJ...691L..49S} {691, L49}

\bibitem[\protect\citeauthoryear{{Syer} \& {Clarke}}{{Syer} \&
  {Clarke}}{1995}]{Syer95}
{Syer} D.,  {Clarke} C.~J.,  1995, \mn@doi [\mnras] {10.1093/mnras/277.3.758},
  \href {https://ui.adsabs.harvard.edu/abs/1995MNRAS.277..758S} {277, 758}

\bibitem[\protect\citeauthoryear{{Tabone}, {Rosotti}, {Lodato}, {Armitage},
  {Cridland}  \& {van Dishoeck}}{{Tabone} et~al.}{2021a}]{Tabone21}
{Tabone} B.,  {Rosotti} G.~P.,  {Lodato} G.,  {Armitage} P.~J.,  {Cridland}
  A.~J.,   {van Dishoeck} E.~F.,  2021a, \mn@doi [\mnras]
  {10.1093/mnrasl/slab124}, \href
  {https://ui.adsabs.harvard.edu/abs/2021MNRAS.tmpL.109T} {}

\bibitem[\protect\citeauthoryear{{Tabone}, {Rosotti}, {Cridland}, {Armitage}
  \& {Lodato}}{{Tabone} et~al.}{2021b}]{Tabone21a}
{Tabone} B.,  {Rosotti} G.~P.,  {Cridland} A.~J.,  {Armitage} P.~J.,   {Lodato}
  G.,  2021b, \mn@doi [\mnras] {10.1093/mnras/stab3442}, \href
  {https://ui.adsabs.harvard.edu/abs/2021MNRAS.tmp.3115T} {}

\bibitem[\protect\citeauthoryear{{Throop} \& {Bally}}{{Throop} \&
  {Bally}}{2005}]{Throop05}
{Throop} H.~B.,  {Bally} J.,  2005, \mn@doi [\apjl] {10.1086/430272}, \href
  {https://ui.adsabs.harvard.edu/abs/2005ApJ...623L.149T} {623, L149}

\bibitem[\protect\citeauthoryear{{Trilling}, {Benz}, {Guillot}, {Lunine},
  {Hubbard}  \& {Burrows}}{{Trilling} et~al.}{1998}]{Trilling98}
{Trilling} D.~E.,  {Benz} W.,  {Guillot} T.,  {Lunine} J.~I.,  {Hubbard} W.~B.,
    {Burrows} A.,  1998, \mn@doi [\apj] {10.1086/305711}, \href
  {https://ui.adsabs.harvard.edu/abs/1998ApJ...500..428T} {500, 428}

\bibitem[\protect\citeauthoryear{{Veras} \& {Armitage}}{{Veras} \&
  {Armitage}}{2004}]{Veras04}
{Veras} D.,  {Armitage} P.~J.,  2004, \mn@doi [\mnras]
  {10.1111/j.1365-2966.2004.07239.x}, \href
  {https://ui.adsabs.harvard.edu/abs/2004MNRAS.347..613V} {347, 613}

\bibitem[\protect\citeauthoryear{Virtanen et~al.,}{Virtanen
  et~al.}{2020}]{Virtanen20}
Virtanen P.,  et~al., 2020, \mn@doi [Nature Methods]
  {10.1038/s41592-019-0686-2}, \href {https://rdcu.be/b08Wh} {17, 261}

\bibitem[\protect\citeauthoryear{{Voelkel}, {Deienno}, {Kretke}  \&
  {Klahr}}{{Voelkel} et~al.}{2021}]{Voelkel21}
{Voelkel} O.,  {Deienno} R.,  {Kretke} K.,   {Klahr} H.,  2021, \mn@doi [\aap]
  {10.1051/0004-6361/202039214}, \href
  {https://ui.adsabs.harvard.edu/abs/2021A&A...645A.131V} {645, A131}

\bibitem[\protect\citeauthoryear{{Winter}, {Clarke}, {Rosotti}, {Ih},
  {Facchini}  \& {Haworth}}{{Winter} et~al.}{2018}]{Win18b}
{Winter} A.~J.,  {Clarke} C.~J.,  {Rosotti} G.,  {Ih} J.,  {Facchini} S.,
  {Haworth} T.~J.,  2018, \mn@doi [\mnras] {10.1093/mnras/sty984}, 478, 2700

\bibitem[\protect\citeauthoryear{{Winter}, {Clarke}, {Rosotti}, {Hacar}  \&
  {Alexander}}{{Winter} et~al.}{2019}]{Winter19b}
{Winter} A.~J.,  {Clarke} C.~J.,  {Rosotti} G.~P.,  {Hacar} A.,   {Alexander}
  R.,  2019, \mn@doi [\mnras] {10.1093/mnras/stz2545}, \href
  {https://ui.adsabs.harvard.edu/abs/2019MNRAS.490.5478W} {490, 5478}

\bibitem[\protect\citeauthoryear{{Winter}, {Kruijssen}, {Chevance}, {Keller}
  \& {Longmore}}{{Winter} et~al.}{2020a}]{Winter20a}
{Winter} A.~J.,  {Kruijssen} J.~M.~D.,  {Chevance} M.,  {Keller} B.~W.,
  {Longmore} S.~N.,  2020a, \mn@doi [\mnras] {10.1093/mnras/stz2747}, \href
  {https://ui.adsabs.harvard.edu/abs/2020MNRAS.491..903W} {491, 903}

\bibitem[\protect\citeauthoryear{{Winter}, {Ansdell}, {Haworth}  \&
  {Kruijssen}}{{Winter} et~al.}{2020b}]{Winter20b}
{Winter} A.~J.,  {Ansdell} M.,  {Haworth} T.~J.,   {Kruijssen} J.~M.~D.,
  2020b, \mn@doi [\mnras] {10.1093/mnrasl/slaa110}, \href
  {https://ui.adsabs.harvard.edu/abs/2020MNRAS.497L..40W} {497, L40}

\bibitem[\protect\citeauthoryear{{Winter}, {Kruijssen}, {Longmore}  \&
  {Chevance}}{{Winter} et~al.}{2020c}]{Winter20c}
{Winter} A.~J.,  {Kruijssen} J.~M.~D.,  {Longmore} S.~N.,   {Chevance} M.,
  2020c, \mn@doi [\nat] {10.1038/s41586-020-2800-0}, \href
  {https://ui.adsabs.harvard.edu/abs/2020Natur.586..528W} {586, 528}

\bibitem[\protect\citeauthoryear{{Wise} \& {Dodson-Robinson}}{{Wise} \&
  {Dodson-Robinson}}{2018}]{Wise18}
{Wise} A.~W.,  {Dodson-Robinson} S.~E.,  2018, \mn@doi [\apj]
  {10.3847/1538-4357/aaaae5}, \href
  {https://ui.adsabs.harvard.edu/abs/2018ApJ...855..145W} {855, 145}

\bibitem[\protect\citeauthoryear{{Youdin} \& {Goodman}}{{Youdin} \&
  {Goodman}}{2005}]{Youdin05}
{Youdin} A.~N.,  {Goodman} J.,  2005, \mn@doi [\apj] {10.1086/426895}, \href
  {https://ui.adsabs.harvard.edu/abs/2005ApJ...620..459Y} {620, 459}

\bibitem[\protect\citeauthoryear{{van Terwisga}, {Hacar}  \& {van
  Dishoeck}}{{van Terwisga} et~al.}{2019}]{vTer19}
{van Terwisga} S.~E.,  {Hacar} A.,   {van Dishoeck} E.~F.,  2019, \mn@doi
  [\aap] {10.1051/0004-6361/201935378}, 628, A85

\makeatother
\end{thebibliography}


\appendix
\section{Dependence on the internal disc wind}
\label{app:int_wind_discuss}
{In this appendix, we consider how variations in our assumed mass-loss rate due to the internally driven wind alter the outcomes of our calculations. Throughout this work, we have adopted a fixed prescription for the mass-loss rate profile driven by a host star of fixed stellar mass. Therefore the variations in outcomes are due to variations in the externally driven wind, which is our focus. However, it has been well-established by several authors that planet growth and migration can be strongly dependent on the internally driven wind \citep{Alexander12b, Ercolano15, Wise18, Monsch21b}. The final outcomes of population synthesis efforts vary with depending on the driving mechanism and resultant mass-loss rate \citep{Jennings18}.}

{To test how sensitive our results are to the choice of wind profile, we here vary the initial wind suppression factor $f_\mathrm{ng}$ for different initial semi-major axes $a_\mathrm{p,0}$. The suppression factor is introduced to obtain disc lifetimes and accretion rate distributions that are comparable to empirical and numerical constraints, as described in Section~\ref{sec:int_wind}. Our choice of $f_\mathrm{ng}=0.1$ and viscous $\alpha=3\times 10^{-3}$ produce reasonable disc life-times and accretion rates. However, alternative physical values of these parameters remain possible, possibly in conjunction with magnetically mediated disc evolution \citep{Tabone21} and dispersal \citep{Bai13}. }

\begin{figure}
    \centering
    \includegraphics[width=\columnwidth]{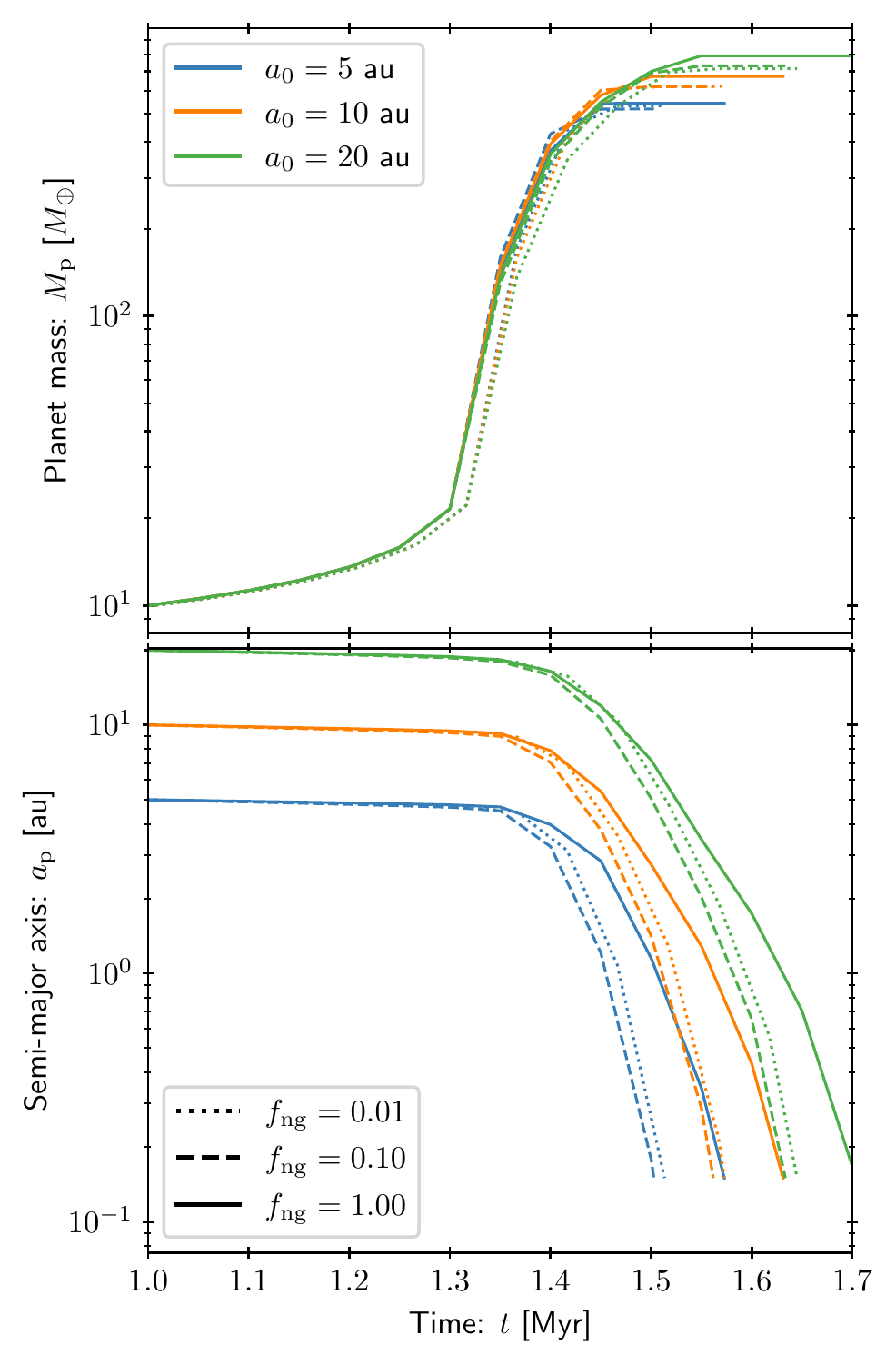}
    \caption{As in Figure~\ref{fig:aM_evol}, but where there is no external wind ($F_\mathrm{FUV} = 0 \, G_0$) and the internal wind suppression factor $f_\mathrm{ng}$ and starting semi-major axis $a_\mathrm{p,0}$ of the planet varies. We show the results of adopting $f_\mathrm{ng}=0.01$, $0.1$~and $1$ as dotted, dashed and solid lines respectively. Semi-major axes $a_\mathrm{p,0}= 5$~au, $10$~au and $20$~au are shown as blue, orange and green respectively. Simulations are stopped when the planet reaches a semi-major axis $a_\mathrm{p}< 0.15$~au.}
    \label{fig:aM_plot_intwind}
\end{figure}

{The results of varying the initial wind suppression factor $f_\mathrm{ng}$ are shown in Figure~\ref{fig:aM_plot_intwind}. We find that increasing the internal wind mass-loss rate does not strongly influence either the growth or the inward migration time-scale of the planet for the chosen parameters. This is in apparent contrast to the findings of previous studies, for which the gap carved by internal photoevaporation results can halt migration, resulting in a dearth of planets close to the wind launching radius \citep[e.g.][]{Jennings18}.}

{We qualitatively compare Figure~\ref{fig:aM_plot_intwind} with the right hand panels of Figure 1 of \citet{Jennings18}. The authors of that study modelled the influence of X-ray, EUV and FUV driven photoevaporation on the separations of giant planets. They adopt a prescription for disc evolution and the migration of planets that is similar to that presented in this work. In all cases (initial conditions and mass-loss rate profiles) explored in both studies, planets migrate inside $0.15$~au on short timescales of a few $0.1$~Myr. \citet{Jennings18} find slower type II migration due to the lower viscous $\alpha \approx 7\times 10^{-4}$, compared to $\alpha=3\cdot 10^{-3}$ in this work. Note that in our work we have also included an (inefficient) type I migration stage. The qualitative trend for type II migration in both studies is similar: increasing the internal photoevaporation rate somewhat decreases the degree of migration. }

{For internal winds alone to produce a dearth of planets at the disc launching radius requires specific parameter choices, which are often poorly empirically constrained \citep{Monsch21b}. It is possible that a range of parameters describing the isolated star-disc system produce such a dearth \citep{Jennings18}. However, the external wind mechanism we have explored in this work offers a complementary way to produce a range of semi-major axes, due to the known variation in the external FUV exposure.}

\bsp	
\label{lastpage}
\end{document}